\documentclass[12pt]{iopart}

\usepackage{iopams}
\usepackage{graphics}
\usepackage{epsfig}

\begin{document}

\title{On the Brownian gas: a field theory with a Poissonian ground state}
\author{Andrea Velenich$^1$, Claudio Chamon$^1$, Leticia F Cugliandolo$^{2}$ and Dirk Kreimer$^{3,4}$}
\address{$^1$ Physics Department, Boston University, 590 Commonwealth Avenue, Boston, MA 02215, U.S.A. \\
$^2$ Universit\'e Pierre et Marie Curie - Paris VI, LPTHE UMR 7589, Place Jussieu, 75252 Paris Cedex 05, France \\
$^3$ Department of Mathematics and Statistics, Boston University, 111 Cummington Street, Boston, MA 02215, U.S.A. \\
$^4$ Institute des Hautes \'Etudes Scientifiques, Le Bois-Marie 35, route de Chartres, F-91440 Bures-sur-Yvette, France}
\eads{\mailto{velenich@bu.edu}}
\begin{abstract}
As a first step towards a successful field theory of Brownian particles in interaction, we study exactly the non-interacting case, its combinatorics and its non-linear time-reversal symmetry.
Even though the particles do not interact, the field theory contains an interaction term: the vertex is the hallmark of the original particle nature of the gas and it enforces the constraint of a strictly positive density field, as opposed to a Gaussian free field.
We compute exactly all the n-point density correlation functions, determine non-perturbatively the Poissonian nature of the ground state and emphasize the futility of any coarse-graining assumption for the derivation of the field theory. We finally verify explicitly, on the n-point functions, the fluctuation-dissipation theorem implied by the time-reversal symmetry of the action.
\end{abstract}

\section*{Introduction}

Glassy features in the dynamical behavior of colloidal suspensions are among the most intriguing aspects of their phenomenology. Their idealization as fluids of interacting Brownian particles might provide the key to understand glassiness for continuous, finite dimensional systems. A satisfying analytic treatment seems, however, still out of reach. \\
The evolution of a system of $N$ Brownian particles can be described either by a system of $N$ stochastic ordinary differential equations or, after introducing a density field, by a stochastic partial differential equation  \cite{K94}, \cite{D96} which in turn can be recast into a field theory.
A field-theoretic approach is the ideal framework for the study of conservation laws, encoded as symmetries of the action. The appeal of a field theory is then twofold: the ease to conceive perturbative series expansions in the (weak) inter-particle potential, and the possibility to obtain non-perturbative results deriving approximate self-consistent equations for the correlators. In either case, the symmetries of the action should act as a guidance and a constraint for any approximation scheme. \\
In order to study the glassy dynamics it is necessary to examine a field-theory in its strongly-interacting regime and, to this aim, attempts have been made to re-sum whole classes of diagrams, at all orders in the interaction potential \cite{MR05}, \cite{ABL06}, \cite{KK06}, \cite{KK07}.
The main obstacle towards a successful resummation scheme is that the time-reversal symmetry of the appropriate action involves non-linear transformations of the fields, relating n-point functions for different n's in an intricate fashion: this precludes the possibility of writing easily-solvable self-consistent closure relations for the n-point functions which respect the symmetry.
In general, a symmetry enforces relations between the field-theoretic correlators; in particular, the time-reversal symmetry of the field theory we are about to study leads to the correct fluctuation-dissipation relations expected between correlation and linear response functions.
The fluctuation-dissipation theorem (FDT) is a firmly grounded physical requirement for an equilibrated system: solutions violating FDT may arise because of terms which drive the system out of equilibrium, explicitly breaking time-reversal symmetry, or as a consequence of the spontaneous symmetry breaking occurring in glassy phases. However, any solution describing the \emph{liquid} phase would clearly be unsatisfactory if it were incompatible with FDT. \\
The aim of this paper is a thorough study of the Brownian gas, which is the $0^{th}$ order term of any perturbative expansion in the inter-particle potential. Here, the challenge posed by the non-linear symmetry is isolated from further complications due to the interactions, the origin of the non-linearity being purely entropic.
The material is organized as follows:
in Section \ref{partform} we define the Brownian gas and compute exactly the space-time dependence of all the n-point correlation functions for the particle density.
In Section \ref{fieldtheo} we briefly review how to formulate a path-integral description of the system and solve the field theory exactly, re-computing all the n-point correlation functions in agreement with the results of Section \ref{partform}. Calculations with field-theoretic methods are non-trivial since the Martin-Siggia-Rose (MSR) action describes an interacting field theory, even though the original system contains non-interacting particles. In Section \ref{poiss} we emphasize the role of the vertex in enforcing the constraint of a strictly positive density field as opposed to a Gaussian free field. No coarse-graining procedure and no cut-off scale are in principle necessary for a proper definition of the theory. A central, non-perturbative result is the proof that the statistics of the ground state of the field theory is Poissonian. In Section \ref{symm} we compute the Jacobian of the non-linear field transformation and show explicitly how the n-point functions computed in the previous sections satisfy the fluctuation-dissipation theorem associated to the time-reversal symmetry of the action.

\section{The particle formalism} \label{partform}

\subsection{Brownian particles}

\noindent In natural units ($k_B=1$ and drag coefficient $\gamma=1$) the equation of motion for a Brownian particle labeled by $j$ moving in $\mathbb{R}^d$ is:
\begin{equation} \label{langevin}
\dot{x}_j(t) = \eta_j(t)
\end{equation}
$\eta_j$ is a Gaussian white noise with probability distribution:
\begin{displaymath}
\mathrm{P}\lbrack \eta_j\rbrack \propto e^{-\frac{1}{4T}\int dt \, \eta_j^2(t)}
\end{displaymath} 
and correlators:
\begin{displaymath} 
\langle \eta_{i,a}(t) \eta_{j,b}(t^{\prime}) \rangle = 2T\delta_{ij}\delta_{ab}\delta(t-t^{\prime})
\end{displaymath}
where $a$ and $b$ label the vector components.
The solution of equation (\ref{langevin}) is:
\begin{equation} \label{sol}
x_j(t) = x_j(0) + \int_0^t dt^{\prime} \eta_j(t^{\prime})
\end{equation}
\noindent Note that for any $j$, $\Delta_j(t) = x_j(t) - x_j(0) = \int_0^t dt^{\prime} \eta_j(t^{\prime})$ is a Gaussian random variable. \\
\\
Soon we will be interested in the value of $\langle (\Delta x_a(t_1) - \Delta x_b(t_2))^2 \rangle_{\eta} \;$, which can be easily computed using:
\begin{eqnarray} \label{Dcorr1}
&& \langle \Delta x_{i,a}(t_1) \Delta x_{j,b}(t_2) \rangle_{\eta} = \int_0^{t_1} dt^{\prime} \int_0^{t_2} dt^{\prime\prime} \; \langle \eta_{i,a}(t^{\prime}) \eta_{j,b}(t^{\prime\prime}) \rangle_{\eta} \nonumber \\
&& = \int_0^{t_1} dt^{\prime} \int_0^{t_2} dt^{\prime\prime} \; 2 T \delta_{ij} \delta_{ab} \delta(t^{\prime}-t^{\prime\prime}) = 2 T \delta_{ij} \delta_{ab} \min(t_1,t_2)
\end{eqnarray}
The result in (\ref{Dcorr1}) is non-vanishing only if $\Delta x$'s relative to the same particle are considered and finally:
\begin{equation} \label{Dcorr2}
\langle (\Delta x_a(t_1) - \Delta x_b(t_2))^2 \rangle_{\eta} =  2T\delta_{ab}|t_1-t_2|
\end{equation}

\subsection{The Brownian gas}

\noindent The Brownian gas consists of $N$ Brownian particles (\ref{langevin}) confined in a $d$-dimensional box of volume $V$ with periodic boundary conditions. 
As usual, the thermodynamic limit is defined as:
\begin{equation} \label{thermlimit}
V \rightarrow \infty \qquad ; \qquad N \rightarrow \infty \qquad ; \qquad \frac{N}{V} \rightarrow \rho_0=\mathrm{const}
\end{equation}
The free parameters describing the gas are its average density $\rho_0$ and its temperature $T$, or equivalently the diffusion coefficient related to the temperature by the Einstein relation: $D=\frac{k_B T}{\gamma}$. In natural units $\gamma=1$ and $k_B=1$, so that $D=T$.  
The natural length-scale in the system is the average separation between particles, defined as $\lambda = \rho_0^{-1/d}$. Also, the well-known relation for diffusive processes $\langle \Delta x^2 \rangle = 2 D\Delta t = 2 T \Delta t$ fixes the natural time-scale of the system to be $\tau = \frac{\lambda^2}{2 T}$. Once distances and time intervals are expressed in these natural units, the Brownian gas becomes a parameter-free universal system.

\subsection{Density correlations}

\noindent The number density field $\varrho(x,t)$ is defined as:

\begin{equation} \label{densfield}
\varrho(x,t) = \sum_{j=1}^N \delta(x-x_j(t)) \qquad \mathrm{or:} \qquad \varrho_k(t) = \frac{1}{V} \sum_{j=1}^N e^{ik\cdot x_j(t)}
\end{equation}
We now compute the correlation of densities measured at $n$ different points in space and time. The average over initial conditions and realizations of the noise is fully deterministic and represents, at least at equilibrium, the expected outcome of a measurement. The average over the initial positions $\langle \ldots \rangle_{I.P.}$ and the realizations of the noise $\langle \ldots \rangle_{\eta}$ are defined in \ref{defconv}. \\
\\
The general structure of the following computations is: 
\begin{itemize}
\item[-] The position of each particle at time $t$ depends both on the initial position and on the realization of the associated noise. Using (\ref{sol}) the two contributions can be separated and the averages over initial positions and noise realizations factorize.
\item[-] The average over initial conditions is computed using formula (\ref{averageIP}). 
\item[-] The average over the noise is performed exploiting the standard relation $\langle e^{-i\psi} \rangle_{\psi} = e^{-\frac{1}{2}\langle \psi^2 \rangle_{\psi}}$, valid for any Gaussian random variable $\psi$ with probability distribution $\mathrm{P}\lbrack \psi\rbrack \propto e^{-\frac{1}{2}\psi^2}$, and applying (\ref{Dcorr1}) and (\ref{Dcorr2}).
\end{itemize}
The final result for a generic connected n-point function is surprisingly compact and is reported in formula (\ref{npt-part}).

\subsubsection{2-point functions}

It is a simple exercise to verify that the 1-point function yields $\rho_0$, as expected from (\ref{thermlimit}). Applying the steps described above, for the 2-point function we obtain:

\begin{displaymath}
\langle \varrho(x_1,t_1)\varrho(x_2,t_2) \rangle_{\eta,I.P.} = 
\sum_{k} \Big( e^{ik\cdot(x_1-x_2)} \frac{N}{V^2} e^{- T k^2 |t_1-t_2|} \Big) + \frac{N(N-1)}{V^2}
\end{displaymath}

\noindent The connected correlation function is:
\begin{eqnarray}
\langle \varrho(x_1,t_1)\varrho(x_2,t_2) \rangle_c &=& \langle \varrho(x_1,t_1)\varrho(x_2,t_2) \rangle_{\eta,I.P.} - \langle \varrho(x_1,t_1) \rangle_{\eta,I.P.} \langle \varrho(x_2,t_2) \rangle_{\eta,I.P.} \nonumber \\
&=& \frac{N}{V^2} \Big( \sum_{k} e^{ik\cdot(x_1-x_2)} e^{- T k^2 |t_1-t_2|} \Big) - \frac{N}{V^2} \nonumber
\end{eqnarray}

\noindent In the continuum limit another factor of $V$ appears in the numerator of the first term
\begin{equation} \label{contlim}
\sum_k = \frac{1}{\Delta k} \sum_k \Delta k \to V \int \frac{d^dk}{(2\pi)^d} 
\end{equation}
which is then finite in the thermodynamic limit whereas the second term vanishes, being suppressed by a factor $\frac{1}{V}$.
In the thermodynamic limit the 2-point function is thus:
\begin{eqnarray}
\langle \varrho(x_1,t_1)\varrho(x_2,t_2) \rangle_c &\to& \rho_0 \int \frac{d^dk}{(2\pi)^d} \; e^{ik\cdot(x_1-x_2)} e^{- T k^2 |t_1-t_2|} \nonumber \\
&=& \rho_0 \frac{1}{(4\pi T |t_1-t_2|)^{d/2}} \exp \Bigg(-\frac{(x_1-x_2)^2}{4 T |t_1-t_2|} \Bigg) \nonumber
\end{eqnarray}

\subsubsection{3-point functions}

The computation of the 3-point function gives some further useful insight.
The generalization to an n-point function is then straightforward.
\begin{eqnarray}
& & \langle \varrho(x_1,t_1)\varrho(x_2,t_2)\varrho(x_3,t_3) \rangle_{\eta,I.P.} \nonumber \\
&& = \sum_{k_1,k_2,k_3} e^{ik_1\cdot x_1} e^{ik_2\cdot x_2} e^{ik_3\cdot x_3} \frac{1}{V^3} \delta_{k_1+k_2+k_3} N \langle e^{-ik_1 \cdot \Delta(t_1)} e^{-ik_2 \cdot \Delta(t_2)} e^{-ik_3 \cdot \Delta(t_3)} \rangle_{\eta} \nonumber \\
& & + \Big( \sum_{k_2,k_3} e^{ik_2\cdot x_2} e^{ik_3\cdot x_3} \frac{1}{V^3} \delta_{k_2+k_3} N(N-1) \langle e^{-ik_2 \cdot \Delta(t_2)} e^{-ik_3 \cdot \Delta(t_3)} \rangle_{\eta} \nonumber \\ 
& & + (1\leftrightarrow 2) + (1\leftrightarrow 3) \Big) + \frac{1}{V^3} N(N-1)(N-2) \nonumber
\end{eqnarray}
Each term in the calculation above is associated to a connected or a disconnected correlation; connected pieces represent single-particle self-correlations to which a formal ``conservation of momentum'' rule is associated by formula (\ref{averageIP}).
Again, the subtraction of disconnected pieces removes, in the thermodynamic limit, all the terms except for the first one, which is the only fully connected term.

\subsubsection{n-point functions}

A connected n-point correlation functions can then be written in the thermodynamic limit as:
\begin{eqnarray} \label{eqc}
& & \langle \varrho(x_1,t_1)\dots\varrho(x_n,t_n) \rangle_c \quad \\
&& = \sum_{k_1,\dots,k_n} e^{ik_1\cdot x_1} \dots e^{ik_n \cdot x_n} \frac{N}{V^n} \delta_{k_1+\dots+k_n} \langle \varrho(k_1,t_1)\dots\varrho(k_n,t_n) \rangle_{\eta,I.P.} \nonumber
\end{eqnarray}
where $\langle \varrho(k_1,t_1)\dots\varrho(k_n,t_n) \rangle_{\eta} = \langle e^{-ik_1 \cdot\Delta(t_1)} \dots e^{-ik_n \cdot\Delta(t_n)} \rangle_{\eta}$. 
After computing the average over the noise, the general n-point correlation function is easily obtained:
\begin{equation} \label{nptk}
\langle \varrho(k_1,t_1)\dots\varrho(k_n,t_n) \rangle_{\eta} = \exp \Big( T\sum_{i<j}^n k_i\cdot k_j |t_i-t_j| \Big)
\end{equation}
Note that $\delta_{k_1+\dots+k_n}$ in (\ref{eqc}) allows to perform easily one of the sums over $k$ and the other $n-1$ summations give, in the continuum approximation, a factor $V^{n-1}$ as in (\ref{contlim}). Hence the whole expression is proportional to $\rho_0$ and it is finite in the thermodynamic limit:
\begin{eqnarray} \label{npt-part}
& & \langle \varrho(x_1,t_1)\dots\varrho(x_n,t_n) \rangle_c \\
&& = \rho_0 \int \frac{d^dk_1}{(2\pi)^d} \ldots \int \frac{d^dk_{n-1}}{(2\pi)^d} \; e^{ik_1\cdot(x_1-x_n)} \ldots e^{ik_{n-1}\cdot(x_{n-1}-x_n)} e^{T\sum_{i<j}^n k_i\cdot k_j |t_i-t_j|} \nonumber
\end{eqnarray}
with the constraint $k_n = -(k_1+\ldots +k_{n-1})$ in the last exponential. 
The space and time-translation invariance of the result are manifest.\\
\\
As we have seen, the only finite contributions in the thermodynamic limit come from single-particle self-correlations. The particles' positions $x_j(t)$ obey the Langevin equation (\ref{langevin}) and the solutions are Gaussian random variables for which all the connected moments $\langle x(t_1)\ldots x(t_n) \rangle_c$ vanish if $n>2$. It could be somewhat surprising to find non-vanishing connected moments of the density for every $n$. The key point is that the density field (\ref{densfield}) is a \emph{function} of the particle positions and a generic function of Gaussian random variables does not have to be, and in this case is not, Gaussianly distributed.

\section{The field-theoretic formalism} \label{fieldtheo}

We now briefly review how to introduce a single differential equation with all the information content of the system of Langevin equations describing the Brownian gas, and how to rephrase it as a field theory.

\subsection{Dean's equation}

The Brownian dynamics of a system of particles can be encoded in a stochastic ordinary differential equation for each particle; since we consider the non-interacting case only, the equations are naturally decoupled:
\begin{displaymath}
\dot{x}_j(t) = \eta_j(t) \qquad ; \qquad j=1,\ldots, N
\end{displaymath}
Defining the density field as in (\ref{densfield}), the system of equations can be formally translated into Dean's equation (\ref{dean-eq}), which is the corresponding partial differential equation governing the evolution of the density field:
\begin{equation} \label{dean-eq}
\partial_t \varrho =  \nabla \cdot \left( \xi\sqrt{\varrho} \right)
+ T \nabla^2 \varrho
\end{equation}
In the process, the $N$ independent noises $\eta_j(t)$ have been turned into a single vector field $\xi(x,t)$ with self-correlation:
\begin{displaymath}
\langle \xi_a(x,t) \xi_b(x^{\prime},t^{\prime}) \rangle = 2T \delta_{ab} \delta(t-t^{\prime}) \delta(x-x^{\prime})
\end{displaymath}
Since in (\ref{dean-eq}) the noise multiplies the density field, the differential equation becomes a meaningful  string of symbols only once a proper interpretation for the noise term is given \cite{VK81}. We adopt the It\^o prescription, which implies a trivial Jacobian \cite{ABL06} for the path-integral in (\ref{genfunc}). \\
\\
It should be stressed that, due to the singular nature of the density field, the natural setting of Dean's equation is a distribution space. In such a setting the equation has been derived exactly, without any coarse-graining assumptions or cut-off scales which are by no means necessary to obtain the differential equation \cite{D96}.

\subsection{The MSR action and the Feynman rules}

\subsubsection{The MSR procedure}

Through a formal procedure first devised by Janssen \cite{J76} following Martin, Siggia and Rose (MSR) \cite{MSR73}, a stochastic differential equation can be recast into a field theory defined by a functional integral. The basic idea is to impose the original equation of motion for the density field $\rho$ by introducing a new field $i\hat{\phi}$ as Lagrange multiplier, whereas the noise is integrated out exploiting its Gaussian measure. \\
\\
After introducing for convenience a shifted density field with a vanishing average \footnote{The spatial average is always well defined, being one of the parameters defining the thermodynamic limit. At equilibrium we expect it to be equivalent to a temporal average at a fixed position and also to an average over the noise histories at fixed position and time.},
\begin{equation} \label{decrho}
\varrho (\vec x,t) = \rho_0+\rho (\vec x,t) \quad \mathrm{s.t.} \quad \langle \varrho(x,t) \rangle = \rho_0 \quad , \quad \langle \rho (\vec x,t) \rangle = 0
\end{equation}
the analytic expression of the MSR generating functional is:
\begin{equation} \label{genfunc}
\mathcal{Z}=\int \mathcal{D}\hat{\phi} \mathcal{D}\rho \; e^{S}
\end{equation}
with action:
\begin{equation} \label{action} 
S=\int \mathrm{d}^dx \, \mathrm{d}t \;  \Big\{ i\hat{\phi}[\partial_t \rho-T\nabla^2\rho] + T\rho_0 [\nabla i\hat{\phi}]^2 + T\rho[\nabla i\hat{\phi}]^2 \Big\}
\end{equation}
The time integration in (\ref{action}) extends from $-\infty$ to $+\infty$ or, equivalently, the initial conditions are shifted to the infinite past: due to the dissipative dynamics, the ground-state of the theory describes then the equilibrated system, forgetful of any initial condition.
We do not specify instead the functional domain over which the path integrals in (\ref{genfunc}) are computed and proceed formally: the functional integrals are regarded as formal series of Gaussian integrals computed over unconstrained (i.e.\ positive and negative) fields whereas, in principle, the integral over $\rho$ should be restricted to sums of $\delta$'s only, as the form of the density field in (\ref{densfield}) would suggest. The exact match of the results we will obtain with the ones calculated in the particle formalism justifies \emph{a posteriori} such an approach, emphasizing the role of the auxiliary field $i\hat{\phi}$ and of the vertex in imposing the equation of motion (\ref{dean-eq}) and enforcing the positivity of the density field.

\subsubsection{Feynman rules} \label{Feynrules} 

For the following analysis it will be convenient to work in a representation involving times $t$ and momenta $k$. Introducing the time-ordering convention to write earlier fields to the right of later fields, the appropriate Feynman rules are then: \\
\begin{center}
\begin{tabular}{l @{\hspace{20 mm}} l} \label{feynrules}
\vspace{4mm}
\includegraphics[width=0.09\textwidth]{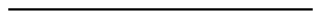} & $\langle \rho\rho \rangle (k,t) = \rho_0 e^{-Tk^2|t|} $ \\
\vspace{4mm}
\includegraphics[width=0.09\textwidth]{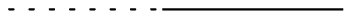} & $\langle i\hat{\phi}\rho \rangle (k,t) = \theta (-t) e^{+Tk^2 t} $ \\
\vspace{4mm}
\includegraphics[width=0.09\textwidth]{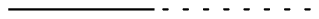} & $\langle \rho i\hat{\phi} \rangle (k,t) = \theta (t) e^{-Tk^2 t} $ \\
\vspace{4mm}
\includegraphics[width=0.09\textwidth]{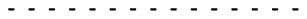} & $\langle i\hat{\phi} i\hat{\phi} \rangle (k,t) = 0 $ \\
\vspace{4mm}
\includegraphics[width=0.09\textwidth]{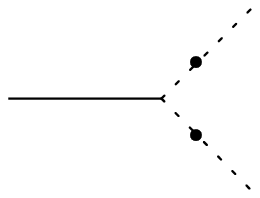} & \raisebox{12pt}{$-2 T k_1\cdot k_2$} \\
\end{tabular}
\end{center}
The dotted edges on the vertex indicate the action of spatial derivatives. All the momenta are considered positive when incoming in the vertex. \\
\\
The propagator $\langle \rho\rho \rangle$ is the density-density correlator and we will refer to the propagators $\langle i\hat{\phi}\rho \rangle$ and $\langle \rho i\hat{\phi} \rangle$ as the advanced and retarded response functions respectively. It is worth emphasizing that such terminology would be appropriate only in the case of Langevin equations with additive noise, for which the occurrence of $i\hat{\phi}$ in a field-theoretic expectation value can be proven to represent the linear response to the variation of a chemical potential $\mu$ coupled to the density field. For the MSR field theory associated to Dean's equation, instead, response functions are characterized not by $i\hat{\phi}$ but by the composite operator $\nabla \cdot ( \varrho \nabla i\hat{\phi} )$ as we will show in Section \ref{symm1}.
Yet, for simplicity, in the following we will refer to the pseudo-response functions $\langle i\hat{\phi}\rho \rangle$ and $\langle \rho i\hat{\phi} \rangle$ as ``response functions'' \emph{tout court}.

\subsubsection{Diagrammatics} \label{diagrammatics}

Although corresponding to a non-interacting particle system, the action (\ref{action}) has a vertex and the field theory is non-Gaussian. The study of the gas is nevertheless easier than the interacting case since all loop diagrams vanish because of causality, as the following example, easily generalizable to arbitrary n-point functions, illustrates:\\
\begin{center}
\includegraphics[width=0.11\textwidth]{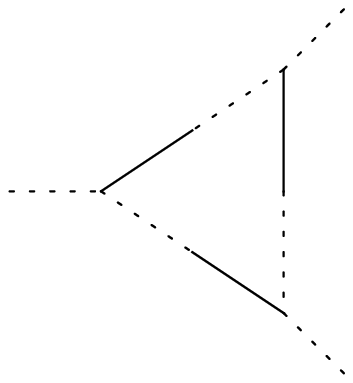}
\begin{picture}(0,0)
\put(-23,48){$t_1$}
\put(-47,17){$t_2$}
\put(-8,8){$t_3$}
\end{picture}
\end{center}
Labeling $t_1$, $t_2$ and $t_3$ the vertices of the triangle, the $\theta$-functions in the propagators impose: $t_1 \leq t_2$, $t_2 \leq t_3$ and $t_3 \leq t_1$. Such constraints can be satisfied only by $t_1=t_2=t_3$, with a vanishing volume in phase space. Hence, only tree diagrams yield non-vanishing amplitudes and such a simplification which will allow to compute exactly all the correlation functions.\\
\\
The structure of the vertex implies that among the edges of a tree there is one and only one correlation-like propagator, all the other propagators being response-like.
The identity below allows to express a correlation-like propagator as a sum of two response-like propagators: 
\begin{eqnarray} \label{prop-dec}
\langle \rho \rho \rangle (k,t) = \rho_0 e^{-T k^2 |t|} &=& \rho_0 \Big[ \theta(t) e^{-T k^2 t} + \theta(-t) e^{T k^2 t} \Big] \nonumber \\
&=& \rho_0  \Big[ \langle \rho i\hat{\phi} \rangle(k,t) + \langle i\hat{\phi} \rho \rangle (k,t) \Big]
\end{eqnarray}
Graphically: $\includegraphics[width=0.07\textwidth]{Feyn1.eps} \; \rightarrow \; \rho_0 ( \; \includegraphics[width=0.07\textwidth]{Feyn3.eps} \; + \; \includegraphics[width=0.07\textwidth]{Feyn2.eps} \; )$. \\
Using relation (\ref{prop-dec}), each diagram can be decomposed into the sum of two simpler diagrams in which all the propagators are response-like: causality, encoded through the $\theta(t)$ functions in the Feynman rules, endows the response-like propagators with a natural orientation so that the concept of ``earlier vertex'' is well-defined.
The above decomposition leads also to the appearance of vertices with three dashed legs whose properties are discussed in Section \ref{c-diag} \\
\\
After applying the decomposition (\ref{prop-dec}) the diagrams contributing to a correlation function can be divided into two subsets:
\begin{itemize} 
 \item[-] c-diagrams: all the external lines are solid lines or, equivalently, one of the vertices has three dashed legs.
 \item[-] r-diagrams: one of the external lines is a dashed line or, equivalently, all the vertices are of the kind depicted in the Feynman rules of Section \ref{Feynrules}.
\end{itemize}
Indicating by $C$ and $R$ the sum of all the c-diagrams and r-diagrams respectively, we have:
\begin{itemize}
\item[-] Response functions = R
\item[-] Correlation functions = C+R 
\end{itemize}
In the next section we will develop a computational scheme for calculating all the diagrams of the theory and, after summing together the appropriate contributions, we will give explicit analytic expressions for the generic n-point density correlations. 
In order to simplify the notation, the temperature will be set to one in the following ($T=1$) an it will be easily reintroduced in the final result (\ref{grand}).

\subsection{Computation of all the r-diagrams}

r-diagrams can be immediately mapped to binary rooted trees by considering the propagator with the external dashed line as the root of the tree. 
Before confronting the general case it is instructive to compute explicitly the first few r-diagrams to better understand and appreciate the combinatorial approach developed in section \ref{comb}. The hasty reader might want to continue there directly. \\
\\
Conventions: \\ 
$t_1<t_2<\ldots <t_{n-1}<t_n$ \\
$m_{ab\ldots yz} := \min{(t_a,t_b,\ldots,t_y,t_z)}$

\subsubsection{3-point}

Applying the Feynman rules and constraining the integration domain according to causality:
\begin{center}
\includegraphics[width=0.08\textwidth]{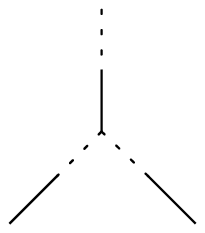}
\begin{picture}(0,0)
\put(-20,37){$1$}
\put(-43,-10){$2$}
\put(-6,-10){$3$}
\end{picture}
\end{center}
\begin{eqnarray} \label{R3}
& & R_3(1,2,3) = \int_{t_1}^{m_{23}} d\bar{t} \; (-2 k_2\cdot k_3) e^{-k_1^2(\bar{t}-t_1)} e^{-k_2^2(t_2-\bar{t})} e^{-k_3^2(t_3-\bar{t})} \\
&& = e^{-k_1^2(m_{23}-t_1)-k_2^2(t_2-m_{23})-k_3^2(t_3-m_{23})} - e^{-k_2^2(t_2-t_1)-k_3^2(t_3-t_1)} \nonumber
\end{eqnarray}
where conservation of momentum: $-k_1^2+k_2^2+k_3^2 = -2 k_2\cdot k_3$ has been used. \\
As explained below, the result can be depicted as:
\begin{equation} \label{R3graphs}
R_3 \quad = \quad \includegraphics[width=0.04\textwidth]{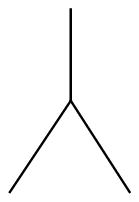} \quad - \quad \includegraphics[width=0.04\textwidth]{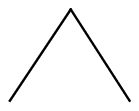}
\begin{picture}(0,0)
\put(-63,23){\footnotesize{$1$}}
\put(-61,11){\footnotesize{$2$}}
\put(-79,-7){\footnotesize{$2$}}
\put(-56,-7){\footnotesize{$3$}}
\put(-7,13){\footnotesize{$1$}}
\put(-22,-7){\footnotesize{$2$}}
\put(-1,-7){\footnotesize{$3$}}
\end{picture}
\end{equation}
In the following we will be careful with the terminology, calling ``diagrams'' the usual Feynman diagrams \cite{BT04},  mapped to integrals by the Feynman rules, and ``graphs'' the objects of the kind drawn in (\ref{R3graphs}), which are a graphical representation of the \emph{result} of such integrals. \\
The crucial property of such a graphical representation is that the group of permutation of labels under which the exponentials in an analytic term are invariant is isomorphic to the group of topologically equivalent relabelings of the associated graph. As we will see starting from the computation of the 4-point r-function, the graphical approach greatly eases the bookkeeping. \\
The rules for drawing a graph given the result of the integrations are very simple:
\begin{itemize} \label{graphrule}
\item[-] To each term of the form $e^{-k_j^2(a-b)}$ associate two vertices labeled `$a$' and `$b$' linked by a line. The edge should be labeled by $k_j^2$, yet we will avoid doing so since the labels for the edges of a graph can be easily guessed considering the momenta labelling the original Feynman diagram and the contractions to obtain the graph.
\item[-] Identify all the vertices with the same label
\end{itemize}
Note that no loops are generated as a consequence of the tree-like structure of the original Feynman diagrams and tadpoles are naturally removed since $a=b$ implies $e^{-k_j^2(a-b)}=1$. \\
Recovering an analytic expression given a labeled graphs is an obvious procedure. \\
Our graphical convention is to orient both, diagrams and graphs, in such a way that the arrow of time points towards the bottom of the page. Then, the label $m_{ab\ldots yz}$ associated to the vertex of a graph is the time labeling the earliest of the underlying leaves. \\

\subsubsection{4-point}

The 4-point r-function can be easily computed from the result of the 3-point r-function: 
\begin{center}
\includegraphics[width=0.12\textwidth]{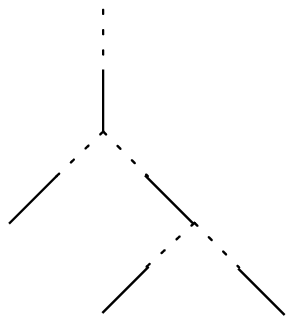}
\begin{picture}(0,0)
\put(-36,56){$1$}
\put(-64,11){$2$}
\put(-32,34){\vector(1,-1){12}}
\put(-24,28){$A$}
\put(-45,-8){$3$}
\put(-5,-8){$4$}
\end{picture}
\end{center}
\begin{eqnarray} \label{R4}
& & r_4(1,2,3,4) = \int_{t_1}^{m_{234}} d\bar{t} \; (+2 k_2\cdot k_A) e^{-k_1^2(\bar{t}-t_1)} e^{-k_2^2(t_2-\bar{t})} R_3(A,3,4) \nonumber \\
&& = e^{-k_1^2(m_{234}-t_1)-k_2^2(t_2-m_{234})-k_A^2(m_{34}-m_{234})-k_3^2(t_3-m_{34})-k_4^2(t_4-m_{34})} \\
& & - e^{-k_2^2(t_2-t_1)-k_A^2(m_{34}-t_1)-k_3^2(t_3-m_{34})-k_4^2(t_4-m_{34})} \nonumber \\
& & - \frac{2 k_2\cdot k_A}{(-k_1^2+k_2^2+k_3^2+k_4^2)} e^{-k_1^2(m_{234}-t_1)-k_2^2(t_2-m_{234})-k_3^2(t_3-m_{234})-k_4^2(t_4-m_{234})} \nonumber \\
& & + \frac{2 k_2\cdot k_A}{(-k_1^2+k_2^2+k_3^2+k_4^2)} e^{-k_2^2(t_2-t_1)-k_3^2(t_3-t_1)-k_4^2(t_4-t_1)} \nonumber
\end{eqnarray}
where $k_A=k_1+k_2=-(k_3+k_4)$.
In the last two terms of the result, rational functions of the external momenta make their first appearance. 
The exponentials and the denominators in the last two lines are independent of the permutation of labels $\{ 2,3,4 \}$ and since the interest is eventually in the sum over all the topologically distinct permutations, we add together the numerators of the rational expressions; due care should be taken since $k_A$ does depend on the particular permutation and the $k_A$'s relative to different permutations need to be distinguished:
\begin{eqnarray} \label{norat4}
& & 2 \big( k_2\cdot k_A + k_3\cdot k_{A^{\prime}} + k_4\cdot k_{A^{\prime\prime}} \big) \nonumber \\
& & = 2 \big( k_2\cdot(k_1+k_2) + k_3\cdot(k_1+k_3) + k_4\cdot(k_1+k_4) \big) \\
& & = 2 \big( k_2^2+k_3^2+k_4^2+k_1(k_2+k_3+k_4) \big) = 2(-k_1^2+k_2^2+k_3^2+k_4^2) \nonumber
\end{eqnarray}
Simplifying the denominators, both rational expressions disappear, leaving a factor of 2. The final result has then the following representation, the capital `R' in $R_4$ representing the sum over all the possible labelings of all the r-diagrams:
\begin{equation} \label{final4}
R_4 \; = \; \sum_{TI} \Big( \; \includegraphics[width=0.05\textwidth]{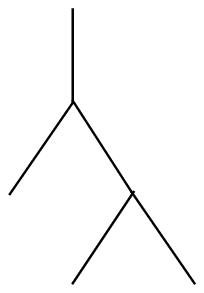} \; - \; \includegraphics[width=0.05\textwidth]{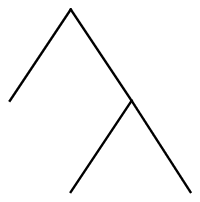} \; - \; 2 \; \includegraphics[width=0.05\textwidth]{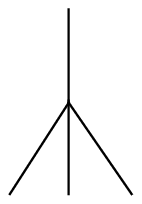} \; + \; 2 \; \includegraphics[width=0.05\textwidth]{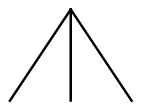} \; \Big)
\end{equation}
$\sum_{TI}$ is a sum over all the topologically inequivalent relabelings of the graphs. Note that the property of two labelings of being TI (topologically inequivalent) or TE (topologically equivalent) depends on the particular graph.
The first two graphs in (\ref{final4}) need to be summed over the 3 TI permutations of external labels $\{ 2,3,4 \}$ (`1' being always fixed since any r-diagram vanishes unless the external dashed line is associated to the earliest time). The last two graphs do not need any further treatment or, equivalently, the 3 possible labelings are topologically equivalent and they have been summed over already to replace the rational functions with the integer `2'.

\subsubsection{5-point} \label{sec5pt}

In this example one last complication, given by the existence of several topologies of binary trees with the same number of external legs, needs to be faced. In order to remove the rational functions it is required to sum not only over the permutations of labels, as before, but also over the possible topologies of trees. Note that in this context diagrams with the same topology but whose propagators have different orientations are considered topologically distinct and they are mapped to different binary rooted trees, namely 5a and 5b below. \\
\\
\emph{Topology a:}
\begin{center}
\includegraphics[width=0.15\textwidth]{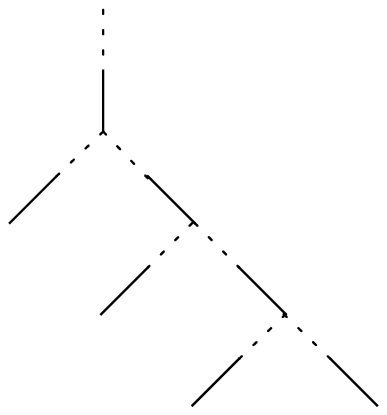}
\begin{picture}(0,0)
\put(-51,67){$1$}
\put(-75,24){$2$}
\put(-48,50){\vector(1,-1){12}}
\put(-42,45){$B$}
\put(-59,8){$3$}
\put(-31,33){\vector(1,-1){12}}
\put(-25,29){$A$}
\put(-43,-9){$4$}
\put(-5,-9){$5$}
\end{picture}
\end{center}

\begin{eqnarray} \label{R5agraphs}
& & r_{5a}(1,2,3,4,5) = \int_{t_1}^{m_{2345}} d\bar{t} \; (+2 k_2\cdot k_B) e^{-k_1^2(\bar{t}-t_1)} e^{-k_2^2(t_2-\bar{t})} r_4(B,3,4,5) = \nonumber \\
\nonumber \\
&& = \quad \includegraphics[width=0.05\textwidth]{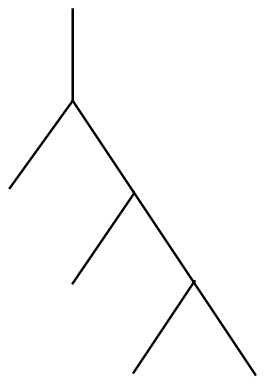}
\begin{picture}(0,0)
\put(-16,32){\footnotesize{$1$}}
\put(-13,22){\footnotesize{$2$}}
\put(-27,9){\footnotesize{$2$}}
\put(-8,15){\footnotesize{$3$}}
\put(-21,0){\footnotesize{$3$}}
\put(-3,7){\footnotesize{$4$}}
\put(-15,-8){\footnotesize{$4$}}
\put(-1,-8){\footnotesize{$5$}}
\end{picture}
\quad - \quad \includegraphics[width=0.05\textwidth]{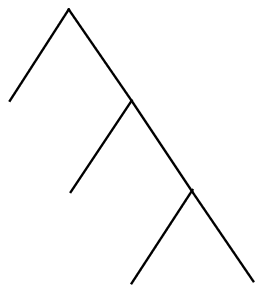}
\begin{picture}(0,0)
\put(-16,25){\footnotesize{$1$}}
\put(-26,9){\footnotesize{$2$}}
\put(-9,16){\footnotesize{$3$}}
\put(-21,1){\footnotesize{$3$}}
\put(-3,7){\footnotesize{$4$}}
\put(-15,-8){\footnotesize{$4$}}
\put(-1,-8){\footnotesize{$5$}}
\end{picture}
\quad - \quad \mathcal{A}_1 \quad \includegraphics[width=0.05\textwidth]{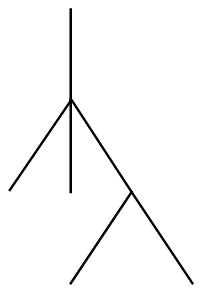}
\begin{picture}(0,0)
\put(-13,31){\footnotesize{$1$}}
\put(-11,20){\footnotesize{$2$}}
\put(-26,4){\footnotesize{$2$}}
\put(-18,4){\footnotesize{$3$}}
\put(-4,9){\footnotesize{$4$}}
\put(-19,-8){\footnotesize{$4$}}
\put(-1,-8){\footnotesize{$5$}}
\end{picture}
\quad + \quad \mathcal{A}_1 \quad \includegraphics[width=0.05\textwidth]{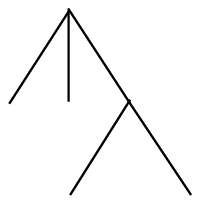} 
\begin{picture}(0,0)
\put(-14,22){\footnotesize{$1$}}
\put(-25,4){\footnotesize{$2$}}
\put(-17,4){\footnotesize{$3$}}
\put(-6,11){\footnotesize{$4$}}
\put(-19,-8){\footnotesize{$4$}}
\put(-1,-8){\footnotesize{$5$}}
\end{picture}
\quad \\
\nonumber \\
&& - \quad \mathcal{A}_2 \quad \includegraphics[width=0.05\textwidth]{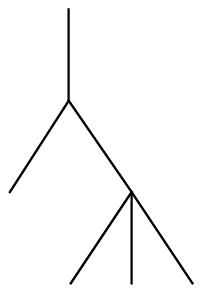} 
\begin{picture}(0,0)
\put(-14,32){\footnotesize{$1$}}
\put(-12,20){\footnotesize{$2$}}
\put(-26,4){\footnotesize{$2$}}
\put(-4,9){\footnotesize{$3$}}
\put(-19,-8){\footnotesize{$3$}}
\put(-10,-8){\footnotesize{$4$}}
\put(-1,-8){\footnotesize{$5$}}
\end{picture}
\quad + \quad \mathcal{A}_2 \quad \includegraphics[width=0.05\textwidth]{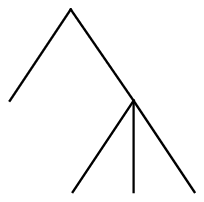} 
\begin{picture}(0,0)
\put(-13,22){\footnotesize{$1$}}
\put(-26,4){\footnotesize{$2$}}
\put(-5,10){\footnotesize{$3$}}
\put(-19,-8){\footnotesize{$3$}}
\put(-10,-8){\footnotesize{$4$}}
\put(-1,-8){\footnotesize{$5$}}
\end{picture}
\quad + \quad \mathcal{A}_3 \quad \includegraphics[width=0.05\textwidth]{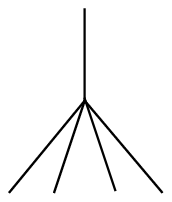} 
\begin{picture}(0,0)
\put(-10,24){\footnotesize{$1$}}
\put(-7,11){\footnotesize{$2$}}
\put(-25,-8){\footnotesize{$2$}}
\put(-17,-8){\footnotesize{$3$}}
\put(-9,-8){\footnotesize{$4$}}
\put(-1,-8){\footnotesize{$5$}}
\end{picture}
\quad - \quad \mathcal{A}_3 \quad \includegraphics[width=0.05\textwidth]{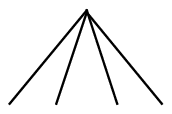}
\begin{picture}(0,0)
\put(-10,14){\footnotesize{$1$}}
\put(-25,-8){\footnotesize{$2$}}
\put(-17,-8){\footnotesize{$3$}}
\put(-9,-8){\footnotesize{$4$}}
\put(-1,-8){\footnotesize{$5$}}
\end{picture}
\nonumber \\
\nonumber
\end{eqnarray}
where:
\begin{eqnarray} \label{rata}
& & \mathcal{A}_1 = \frac{2 k_2 \cdot k_B}{-k_1^2+k_2^2+k_3^2+k_A^2} \qquad
\mathcal{A}_2 = \frac{2 k_3 \cdot k_A}{-k_B^2+k_3^2+k_4^2+k_5^2} \nonumber \\
& & \mathcal{A}_3 = \mathcal{A}_2 \frac{2 k_2 \cdot k_B}{-k_1^2+k_2^2+k_3^2+k_4^2+k_5^2}
\end{eqnarray}
\emph{Topology b:}
\begin{center}
\includegraphics[width=0.15\textwidth]{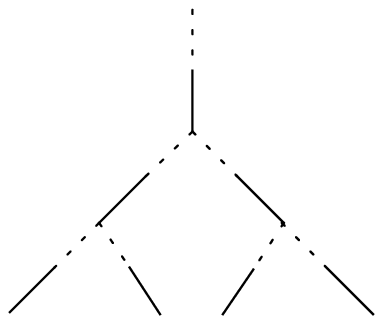}
\begin{picture}(0,0)
\put(-36,53){$1$}
\put(-75,-9){$2$}
\put(-45,33){\vector(-1,-1){12}}
\put(-62,27){$C$}
\put(-45,-9){$3$}
\put(-30,33){\vector(1,-1){12}}
\put(-20,27){$A$}
\put(-34,-9){$4$}
\put(-4,-9){$5$}
\end{picture}
\end{center}

\begin{eqnarray} \label{R5bgraphs}
& & r_{5b}(5,4,3,2,1) = \int_{t_1}^{m_{2345}} d\bar{t} (-2 k_A\cdot k_C) e^{-k_1^2(\bar{t}-t_1)} R_3(C,2,3) R_3(A,4,5) \nonumber \\
\nonumber \\
&& = \quad \includegraphics[width=0.06\textwidth]{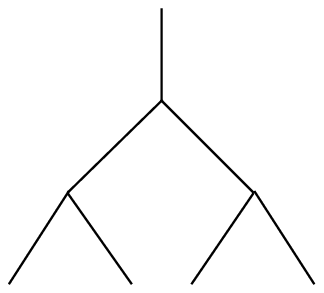} 
\begin{picture}(0,0)
\put(-12,24){\footnotesize{$1$}}
\put(-9,14){\footnotesize{$2$}}
\put(-29,6){\footnotesize{$2$}}
\put(-2,6){\footnotesize{$4$}}
\put(-30,-8){\footnotesize{$2$}}
\put(-19,-8){\footnotesize{$3$}}
\put(-13,-8){\footnotesize{$4$}}
\put(-3,-8){\footnotesize{$5$}}
\end{picture}
\quad - \quad \includegraphics[width=0.06\textwidth]{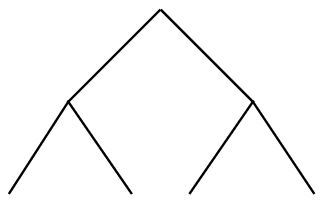} 
\begin{picture}(0,0)
\put(-11,16){\footnotesize{$1$}}
\put(-29,6){\footnotesize{$2$}}
\put(-2,6){\footnotesize{$4$}}
\put(-30,-8){\footnotesize{$2$}}
\put(-19,-8){\footnotesize{$3$}}
\put(-13,-8){\footnotesize{$4$}}
\put(-3,-8){\footnotesize{$5$}}
\end{picture}
\quad - \quad \mathcal{B}_1 \quad \includegraphics[width=0.05\textwidth]{bt5_3.eps} 
\begin{picture}(0,0)
\put(-14,32){\footnotesize{$1$}}
\put(-12,20){\footnotesize{$2$}}
\put(-27,4){\footnotesize{$2$}}
\put(-18,4){\footnotesize{$3$}}
\put(-6,10){\footnotesize{$4$}}
\put(-19,-8){\footnotesize{$4$}}
\put(-1,-8){\footnotesize{$5$}}
\end{picture}
\quad + \quad \mathcal{B}_1 \quad \includegraphics[width=0.05\textwidth]{bt5_4.eps} 
\begin{picture}(0,0)
\put(-13,22){\footnotesize{$1$}}
\put(-27,4){\footnotesize{$2$}}
\put(-18,4){\footnotesize{$3$}}
\put(-6,11){\footnotesize{$4$}}
\put(-19,-8){\footnotesize{$4$}}
\put(-1,-8){\footnotesize{$5$}}
\end{picture}
\quad \\
\nonumber \\
&& - \quad \mathcal{B}_2 \quad \includegraphics[width=0.05\textwidth]{bt5_3.eps} 
\begin{picture}(0,0)
\put(-14,32){\footnotesize{$1$}}
\put(-12,20){\footnotesize{$2$}}
\put(-27,4){\footnotesize{$4$}}
\put(-18,4){\footnotesize{$5$}}
\put(-5,9){\footnotesize{$2$}}
\put(-18,-7){\footnotesize{$2$}}
\put(-1,-7){\footnotesize{$3$}}
\end{picture}
\quad + \quad \mathcal{B}_2 \quad \includegraphics[width=0.05\textwidth]{bt5_4.eps} 
\begin{picture}(0,0)
\put(-13,22){\footnotesize{$1$}}
\put(-27,4){\footnotesize{$4$}}
\put(-18,4){\footnotesize{$5$}}
\put(-5,9){\footnotesize{$2$}}
\put(-18,-7){\footnotesize{$2$}}
\put(-1,-7){\footnotesize{$3$}}
\end{picture}
\quad + \quad \mathcal{B}_3 \quad \includegraphics[width=0.05\textwidth]{bt5_7.eps} 
\begin{picture}(0,0)
\put(-10,24){\footnotesize{$1$}}
\put(-8,12){\footnotesize{$2$}}
\put(-26,-7){\footnotesize{$2$}}
\put(-18,-7){\footnotesize{$3$}}
\put(-10,-7){\footnotesize{$4$}}
\put(-1,-7){\footnotesize{$5$}}
\end{picture}
\quad - \quad \mathcal{B}_3 \quad \includegraphics[width=0.05\textwidth]{bt5_8.eps}
\begin{picture}(0,0)
\put(-10,14){\footnotesize{$1$}}
\put(-26,-7){\footnotesize{$2$}}
\put(-18,-7){\footnotesize{$3$}}
\put(-10,-7){\footnotesize{$4$}}
\put(-1,-7){\footnotesize{$5$}}
\end{picture}
\nonumber \\
\nonumber
\end{eqnarray}
where:
\begin{eqnarray} \label{ratb}
& & \mathcal{B}_1 = \frac{-2 k_A \cdot k_C}{-k_1^2+k_2^2+k_3^2+k_A^2} \qquad
\mathcal{B}_2 = \frac{-2 k_A \cdot k_C}{-k_1^2+k_4^2+k_5^2+k_C^2} \nonumber \\
& & \mathcal{B}_3 = \frac{-2 k_A \cdot k_C}{-k_1^2+k_2^2+k_3^2+k_4^2+k_5^2}
\end{eqnarray}
In the following, with $5_{a,n^{th}}$ and $5_{b,n^{th}}$ we will refer to the $n^{th}$ graph appearing in (\ref{R5agraphs}) and (\ref{R5bgraphs}) respectively. \\
\\
\textbf{A subtlety:} \\ 
The group $\mathcal{G}_{5a}$ of TI relabelings for Feynman diagram 5a contains 12 elements and induces 12 relabelings for the graphs in (\ref{R5agraphs}). For each graph in (\ref{R5agraphs}) we can factorize $\mathcal{G}_{5a}$ as a product of TI and TE relabelings, the factorization depending on the particular graph considered. The factorization is possible since the graphs generated by contracting edges in diagram 5a have a higher symmetry than the diagram itself. Unfortunately this is not always the case; in fact, Feynman diagram 5b has only 3 TI relabelings. However some of the graphs in (\ref{R5bgraphs}), namely $5_{b,3^{rd}}, 5_{b,4^{th}}, 5_{b,5^{rd}}$ and $5_{b,6^{th}}$, admit 6 TI relabelings since the edge contractions have lowered the symmetry of 5b. Nevertheless, the troublesome graphs appear in pairs ($\{5_{b,3^{rd}}, 5_{b,5^{th}} \}$ and $\{5_{b,4^{rd}}, 5_{b,6^{th}} \}$) so that $\mathcal{G}_{5b}$ induces 3 relabelings on each of them for a total of 6 relabelings associated to each topology. The 3 elements in $\mathcal{G}_{5b}$ are thus enough to induce all the TI relabelings for any graph in (\ref{R5bgraphs}). In general, whenever the contractions reduce the symmetry factor of a diagram by a factor $n$, $n$ copies of the same graph appear, so that the TI labelings of the original diagram are sufficient to induce all the possible TI labelings for the topology of the graph. This is a general phenomenon deriving from the fact that the graphs are decorations of the underlying tree structure determined by the Feynman diagrams. \\
\\
The results for diagrams 5a and 5b ((\ref{R5agraphs}) and (\ref{R5bgraphs}) respectively) contain several graphs with the same topology: examples of such terms are $\{ 5_{a,3^{rd}}, 5_{b,3^{rd}}, 5_{b,5^{th}} \}$, $\{ 5_{a,4^{th}}, 5_{b,4^{th}}, 5_{b,6^{th}} \}$, $\{ 5_{a,7^{th}}, 5_{b,7^{th}} \}$ and $\{ 5_{a,8^{th}}, 5_{b,8^{th}} \}$. A labeled graph fully determines the exponentials of the analytic expressions, which are invariant under the TE relabelings of the graph. Instead, the rational functions in (\ref{rata}) and (\ref{ratb}) do depend on the labeling. As for the 4-point function in (\ref{norat4}), by summing over all the TE relabelings for each graph we now replace the rational functions in (\ref{R5agraphs}) and (\ref{R5bgraphs}) with integer numbers. 
The computations are straightforward and can be summarized as follows: for the topology of the groups $\{ 5_{a,3^{rd}}, 5_{b,3^{rd}}, 5_{b,5^{th}} \}$ and $\{ 5_{a,4^{th}}, 5_{b,4^{th}}, 5_{b,6^{th}} \}$ the Feynman diagram 5a induces 12 labelings and 5b induces 6. The sum over the 18 labelings yields a factor of 2. For the topology of the groups $\{ 5_{a,7^{th}}, 5_{b,7^{th}} \}$ and $\{ 5_{a,8^{th}}, 5_{b,8^{th}} \}$ the sum over the 12+3 induced labelings yields a factor of 6.
Finally, the result can be expressed in terms of a linear combination of graphs with integer coefficients, summed over TI relabelings only:
\begin{eqnarray} \label{R5}
&& R_5 \; = \; R_{5a} + R_{5b} \; = \; \sum_{TI} \Big( \; \includegraphics[width=0.05\textwidth]{bt5_1.eps} \; - \; \includegraphics[width=0.05\textwidth]{bt5_2.eps} \; + \; \includegraphics[width=0.06\textwidth]{bt5_9.eps} \; - \; \includegraphics[width=0.06\textwidth]{bt5_10.eps} \; \\
&& - \; 2 \; \includegraphics[width=0.05\textwidth]{bt5_3.eps} \; + \; 2 \; \includegraphics[width=0.05\textwidth]{bt5_4.eps} \; - \; 2 \; \includegraphics[width=0.05\textwidth]{bt5_5.eps} \; + \; 2 \; \includegraphics[width=0.05\textwidth]{bt5_6.eps} \; + \; 6 \; \includegraphics[width=0.05\textwidth]{bt5_7.eps} \; - \; 6 \; \includegraphics[width=0.05\textwidth]{bt5_8.eps} \; \Big) \nonumber
\end{eqnarray}

\subsubsection{n-point} \label{comb}

The previous sections clearly show that the amount of combinatorics required to compute higher order r-functions rapidly becomes prohibitive and we need to automate the computation of n-point diagrams. This can be achieved in two steps: first, starting from a Feynman diagram, generate all the topologies of the relevant graphs, and second, associate to each graph the appropriate combinatorial integer factor.
Indeed, given an n-point Feynman r-diagram, $X$, there is no need to apply the Feynman rules, perform explicitly the computation and then draw the corresponding graphs. Instead, the graphs representing the result are readily obtained by the following procedure:
\begin{itemize}
\item[-] Consider all the graphs drawn after contracting $m$ ($m=0,\ldots,n-2$) among the contractible edges of $X$, where a ``contractible edge'' is defined to be either an internal edge or the external edge labeled by the earliest time (which in the r-graphs is always drawn as the uppermost edge, with the appearance a stalk). 
\item[-] Multiply each graph by a factor $(-1)^m$. 
\item[-] To each vertex with $l$ dangling lines associate a factor $\mathcal{R}_l=(l-1)!$. The integer factor associated to the graph is then the product of all the coefficients associated to its vertices. When considering an explicit labeling, the label associated to a branching is the smallest (earliest) label of its descendant leaves.
\item[-] Sum each graph over its topologically inequivalent (TI) labelings.
\end{itemize}
The first two rules follow naturally from an iterative application of the fundamental theorem of calculus: $\int_a^b dt f(t) = F(b)-F(a)$. In fact, an integral over time is associated to each vertex of a Feynman diagram and the evaluation of the primitive at the integration boundaries determines the sign and the survival or the contraction of the edge in a graph. The third prescription depends on the specific form of the Feynman rules and its proof is reported in \ref{factorial}. All the above prescriptions should be familiar after working out explicitly the analytic computation of the 3, 4, and 5-point functions of the previous sections. \\
\\
The next section aims at motivating one further prescription and at showing explicitly the remaining steps for the calculation of density correlation functions. Some readers might want to go directly to Section \ref{goto}.

\subsection{Computation of all the c-diagrams} \label{c-diag}

In \ref{diagrammatics} we have defined the c-diagrams and noticed the appearance of a new vertex with three dashed legs which was not explicitly present in the action (\ref{action}). We recall that the new vertex is a consequence of the application of the decomposition (\ref{prop-dec}). After summing over the three possible ways of orienting the original r-vertices, or, equivalently, of distributing derivatives marked as dots on the edges, it is convenient to introduce such vertex, called c-vertex, as a new effective Feynman rule:
\begin{displaymath} 
\includegraphics[width=0.05\textwidth]{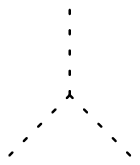} \quad = \quad \includegraphics[width=0.05\textwidth]{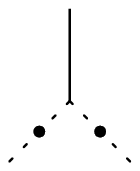} \quad + \quad \includegraphics[width=0.05\textwidth]{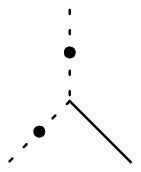} \quad + \quad \includegraphics[width=0.05\textwidth]{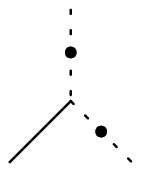}
\end{displaymath}
with analytic expression:
\begin{equation} \label{c-vertex}
-2 k_1\cdot k_2 -2 k_1\cdot k_3 -2 k_2\cdot k_3 = 2 k_1^2 + 2 k_2^2 + 2 k_1\cdot k_2 = k_1^2 + k_2^2 + k_3^2 
\end{equation}
It is easy to prove that the Feynman diagrams generated with the original Feynman rules of section (\ref{feynrules}) and the decomposition (\ref{prop-dec}) are the same as the ones generated with a new set of Feynman rules consisting of an r-vertex, a c-vertex and response-like propagators only. Diagrams generated only with r-vertices are r-diagrams; the ones in which one c-vertex is present are c-diagrams. \\
The application of the Feynman rules is straightforward and the computations of the integrals very similar to the ones for the r-functions. We just report some results, emphasizing the differences with the previous calculations.

\subsubsection{3-point}
\begin{center}
\includegraphics[width=0.10\textwidth]{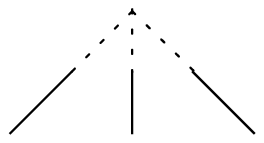}
\begin{picture}(0,0)
\put(-53,-9){$1$}
\put(-29,-9){$2$}
\put(-5,-9){$3$}
\end{picture}
\end{center}
\begin{eqnarray} \label{C3}
& & c_3(1,2,3) = \int_{-\infty}^{m_{123}} d\bar{t} \; (k_1^2+k_2^2+k_3^2) e^{-k_1^2(t_1-\bar{t})} e^{-k_2^2(t_2-\bar{t})} e^{-k_3^2(t_3-\bar{t})} \nonumber \\
&& = e^{-k_1^2(t_1-m_{123})-k_2^2(t_2-m_{123})-k_3^2(t_3-m_{123})} 
\end{eqnarray}
Note the different integration domain in comparison with (\ref{R3}).
The graphical representation of (\ref{C3}) is:
\begin{displaymath}
C_3(1,2,3) \quad = \quad \includegraphics[width=0.05\textwidth]{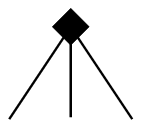}
\begin{picture}(0,0)
\put(-7,14){\footnotesize{$1$}}
\put(-25,-8){\footnotesize{$1$}}
\put(-14,-8){\footnotesize{$2$}}
\put(-3,-8){\footnotesize{$3$}}
\end{picture}
\end{displaymath}
The $\includegraphics[width=0.010\textheight]{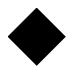}$ which marks the c-vertex is just a bookkeeping symbol whose importance will be clear soon. \\
\\
We are eventually interested in the computation of the 3-point connected density correlation function which is the sum of the 3-point r-function and c-function. Indicating as usual with capital letters the sum over topologically inequivalent relabelings (in this case just one), the graphical result is:
\begin{eqnarray}
\fl
&& \langle \varrho(k_1,t_1)\varrho(k_2,t_2)\varrho(k_3,t_3) \rangle_{c} \; = \; R_3 + C_3 \; = \; \Big( \;\, \includegraphics[width=0.04\textwidth]{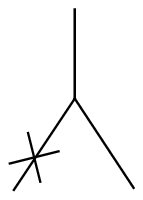}
\begin{picture}(0,0)
\put(-7,22){\footnotesize{$1$}}
\put(-6,11){\footnotesize{$2$}}
\put(-20,-7){\footnotesize{$2$}}
\put(-2,-7){\footnotesize{$3$}}
\end{picture} \; - \; \includegraphics[width=0.04\textwidth]{bt3_2.eps}
\begin{picture}(0,0)
\put(-7,13){\footnotesize{$1$}}
\put(-22,-7){\footnotesize{$2$}}
\put(-3,-7){\footnotesize{$3$}}
\end{picture}
 \;\, \Big) \; + \; \Big( \;\, \includegraphics[width=0.04\textwidth]{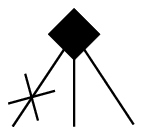} 
\begin{picture}(0,0)
\put(-5,11){\footnotesize{$1$}}
\put(-21,-8){\footnotesize{$1$}}
\put(-11,-8){\footnotesize{$2$}}
\put(-2,-8){\footnotesize{$3$}}
\end{picture}
\;\; \Big) \; = \; \includegraphics[width=0.007\textheight]{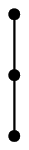}
\begin{picture}(0,0)
\put(0,27){\footnotesize{$1$}}
\put(0,14){\footnotesize{$2$}}
\put(0,0){\footnotesize{$3$}}
\end{picture}
\nonumber
\end{eqnarray}
where we have crossed the edges to be removed when applying the contraction rule for edges with the same label, as described in section \ref{graphrule}. Finally, the last two graphs cancel each other as can be explicitly noted comparing (\ref{R3}) and (\ref{C3}), the $\includegraphics[width=0.010\textheight]{diamond.eps}$ being irrelevant in this context. In the result, the vertices of the ladder graph are pinpointed for clarity by black dots.

\subsubsection{4-point}

Other n-point functions with $n>3$ can be generated recursively from one c-vertex and appropriate combinations of r-functions. For the 4-point c-function we have:
\begin{center}
\includegraphics[width=0.14\textwidth]{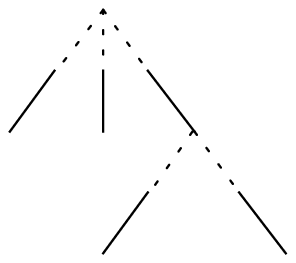}
\begin{picture}(0,0)
\put(-70,18){$1$}
\put(-48,18){$2$}
\put(-36,49){\vector(3,-4){12}}
\put(-28,42){$A$}
\put(-50,-9){$3$}
\put(-6,-9){$4$}
\end{picture}
\end{center}
\begin{eqnarray} \label{c4pt}
& & c_4(1,2,3,4) = \int_{-\infty}^{m_{1234}} d\bar{t} \; (k_1^2+k_2^2+k_A^2) e^{-k_1^2(t_1-\bar{t})} e^{-k_2^2(t_2-\bar{t})} r_3(A,3,4) \nonumber \\
&& = e^{-k_1^2(t_1-m_{1234})-k_2^2(t_2-m_{1234})-k_A^2(m_{34}-m_{1234})-k_3^2(t_3-m_{34})-k_4^2(t_4-m_{34})} \\
& & - \frac{k_1^2+k_2^2+k_A^2}{(k_1^2+k_2^2+k_3^2+k_4^2)} e^{-k_1^2(t_1-m_{1234})-k_2^2(t_2-m_{1234})-k_3^2(t_3-m_{1234})-k_4^2(t_4-m_{1234})} \nonumber
\end{eqnarray}
or, graphically:
\begin{eqnarray} \label{c4ptgr}
c_4(1,2,3,4) \quad = \quad \includegraphics[width=0.05\textwidth]{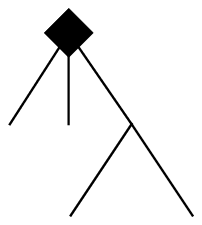}
\begin{picture}(0,0)
\put(-11,20){\footnotesize{$1$}}
\put(-26,4){\footnotesize{$2$}}
\put(-17,4){\footnotesize{$3$}}
\put(-5,10){\footnotesize{$4$}}
\put(-18,-8){\footnotesize{$4$}}
\put(-1,-8){\footnotesize{$5$}}
\end{picture} 
\quad - \quad \mathcal{C} \quad \includegraphics[width=0.05\textwidth]{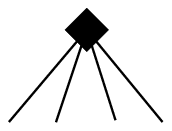}
\begin{picture}(0,0)
\put(-7,11){\footnotesize{$1$}}
\put(-27,-8){\footnotesize{$2$}}
\put(-18,-8){\footnotesize{$3$}}
\put(-10,-8){\footnotesize{$4$}}
\put(-1,-8){\footnotesize{$5$}}
\end{picture}
\end{eqnarray}
where:
\begin{equation}
\mathcal{C} = \frac{k_1^2+k_2^2+k_A^2}{(k_1^2+k_2^2+k_3^2+k_4^2)}
\nonumber
\end{equation}
The Feynman diagram admits 6 TI permutations: differently from the case of the 4-point r-function, the allowed permutations for the 4-point c-function involve all the labels $\{ 1,2,3,4 \}$. The exponential in the second term of (\ref{c4pt}) is invariant under such relabelings (or, equivalently, the relabelings are TE for the second graph in (\ref{c4ptgr})). Since the denominator of $\mathcal{C}$ is also invariant, we can just add together the numerators corresponding to different labelings. Rewriting $(k_1^2+k_2^2+k_A^2)$ as $(2 k_1^2+2 k_2^2+ 2 k_1\cdot k_2)$ and summing over the 6 relabelings:
\begin{eqnarray}
&& (2 k_3^2+2 k_4^2+ 2 k_3\cdot k_4) + (2 k_1^2+2 k_2^2+ 2 k_1\cdot k_2) + (2 k_1^2+2 k_3^2+ 2 k_1\cdot k_3) + \nonumber \\ 
&& + (2 k_1^2+2 k_4^2+ 2 k_1\cdot k_4) + (2 k_2^2+2 k_3^2+ 2 k_2\cdot k_3) + (2 k_2^2+2 k_4^2+ 2 k_2\cdot k_4) = \nonumber \\
&& = 5(k_1^2+k_2^2+k_3^2+k_4^2) \nonumber
\end{eqnarray}
After the sum, the rational function is replaced by a factor of 5. It is clear that the new vertex $\includegraphics[width=0.010\textheight]{diamond.eps}$ generates combinatorial factors different from what we calculated for the r-functions and this justifies a new graphical symbol for it. \\
The final result for the 4-point c-function is:
\begin{equation} \label{finalC4}
C_4 \; = \; \sum_{TI} \Big( \; \includegraphics[width=0.05\textwidth]{gc4_1.eps} \;  - \; 5 \; \includegraphics[width=0.05\textwidth]{gc4_2.eps} \; \Big)
\end{equation}
In order to obtain the 4-point connected density correlation function we add together (\ref{final4}) and (\ref{finalC4}):
\begin{eqnarray}
&& \langle \varrho(k_1,t_1)\varrho(k_2,t_2)\varrho(k_3,t_3)\varrho(k_4,t_4) \rangle_{c} \; = \; \sum_{TI} \Bigg[ \Big( \; \includegraphics[width=0.05\textwidth]{bt4_1.eps} \; - \; \includegraphics[width=0.05\textwidth]{bt4_2.eps} \; \nonumber \\
&& - \; 2 \; \includegraphics[width=0.05\textwidth]{bt4_3.eps} \; + \; 2 \; \includegraphics[width=0.05\textwidth]{bt4_4.eps} \; \Big) \; + \;
\Big( \; \includegraphics[width=0.05\textwidth]{gc4_1.eps} \;  - \; 5 \; \includegraphics[width=0.05\textwidth]{gc4_2.eps} \; \Big) \Bigg] \; = \; \includegraphics[width=0.008\textwidth]{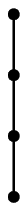} \nonumber
\begin{picture}(0,0)
\put(0,29){\footnotesize{$1$}}
\put(0,19){\footnotesize{$2$}}
\put(0,9){\footnotesize{$3$}}
\put(0,-1){\footnotesize{$4$}}
\end{picture}
\end{eqnarray}
The reader is invited to perform the simple sum over the TI relabelings and verify explicitly that, the result for the density correlator is again a ladder graph.

\subsubsection{5-point}

The general mechanism should be clear by now. As before, for the 5-point functions two topologies are present. We only report the final result, after summing over the two topologies and all the TE relabelings for each graph:
\begin{eqnarray} \label{C5}
\fl
& & C \; = \; C_{5a} + C_{5b} \; = \; \sum_{TI} \Bigg[ \;\, \includegraphics[width=0.05\textwidth]{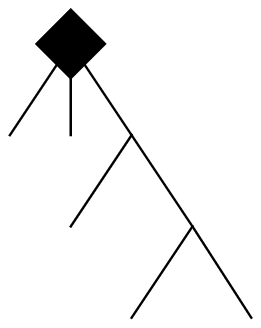} \; + \; \includegraphics[width=0.07\textwidth]{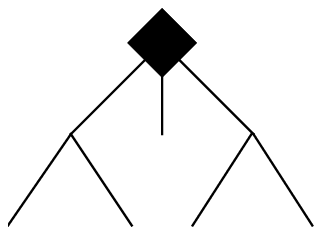} \; - \; 5 \; \includegraphics[width=0.06\textwidth]{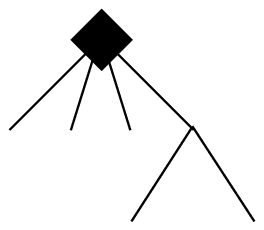} \; - \; 2 \; \includegraphics[width=0.05\textwidth]{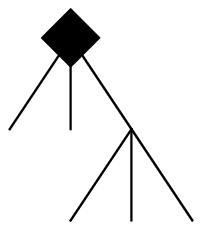} \; + \; 26 \; \includegraphics[width=0.05\textwidth]{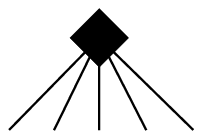} \; \Bigg]
\end{eqnarray}
Note that the factor of 2 associated to the fourth term of the r.h.s.\ is a $(3-1)!$ from the sub-tree with 3 dangling lines. Moreover, an interesting pattern $1,5,26,\ldots$ appears for the integers associated to a c-vertex with $l$ dangling lines. In \ref{stirling} it will be identified with the sequence of generalized Stirling numbers (Sloane A001705, \cite{OEIS}). \\
Verifying that the sum of (\ref{R5}) and (\ref{C5}) yields the 5-vertex ladder graph is a lengthy but rewarding exercise. In \ref{ladders} it is proven for general n-point functions that such ladder graphs are the graphical representation of the very same analytic result obtained in (\ref{nptk}) with the particle formalism.

\subsubsection{n-point} \label{goto}

The general recipe of section \ref{comb} for generating graphs with their integer coefficient can now be complemented with a new prescription:
\begin{itemize}
\item[-] To each c-vertex with $l$ dangling lines associate a factor $\mathcal{C}_l=S_{l-2}$, where $S_l$ is the $l^{th}$ generalized Stirling number (see \ref{stirling} for a proof). The c-vertex is labeled by `1', being always the earliest vertex in a Feynman diagram.
\end{itemize}

\subsection{Graph cancellations}

Up to this point we have computed all the r-diagrams and all the c-diagrams writing them as sums over topologically inequivalent labeling of graphs. At least in the simplest examples, after combining them to compute an n-point density correlation function, we have shown that the result is extremely simple, namely a ladder graph, i.e. a single time-ordered labeling of a single graph, as displayed in (\ref{nladders}).
The equivalence of the analytic expression corresponding to the ladder graphs and the result for the n-point density correlation function computed with the particle formalism in (\ref{nptk}) is explicitly shown in \ref{ladders}.
\begin{eqnarray} \label{nladders}
\includegraphics[width=0.04\textwidth]{bt3_1.eps}
\begin{picture}(0,0)
\put(-7,24){\footnotesize{$1$}}
\put(-5,10){\footnotesize{$2$}}
\put(-21,-8){\footnotesize{$2$}}
\put(-1,-8){\footnotesize{$3$}}
\end{picture}
\; \rightarrow \; \includegraphics[width=0.009\textheight]{ladder3.eps}
\begin{picture}(0,0)
\put(0,35){\footnotesize{$1$}}
\put(0,17){\footnotesize{$2$}}
\put(0,0){\footnotesize{$3$}}
\end{picture}
\qquad ; \qquad \includegraphics[width=0.05\textwidth]{bt4_1.eps} 
\begin{picture}(0,0)
\put(-14,33){\footnotesize{$1$}}
\put(-12,19){\footnotesize{$2$}}
\put(-27,4){\footnotesize{$2$}}
\put(-5,9){\footnotesize{$3$}}
\put(-19,-7){\footnotesize{$3$}}
\put(-1,-7){\footnotesize{$4$}}
\end{picture}
\; \rightarrow \; \includegraphics[width=0.007\textheight]{ladder4.eps}
\begin{picture}(0,0)
\put(0,39){\footnotesize{$1$}}
\put(0,26){\footnotesize{$2$}}
\put(0,12){\footnotesize{$3$}}
\put(0,-1){\footnotesize{$4$}}
\end{picture}
\qquad ; \qquad \includegraphics[width=0.05\textwidth]{bt5_1.eps} 
\begin{picture}(0,0)
\put(-17,31){\footnotesize{$1$}}
\put(-14,22){\footnotesize{$2$}}
\put(-26,9){\footnotesize{$2$}}
\put(-9,14){\footnotesize{$3$}}
\put(-21,0){\footnotesize{$3$}}
\put(-4,6){\footnotesize{$4$}}
\put(-15,-8){\footnotesize{$4$}}
\put(-1,-8){\footnotesize{$5$}}
\end{picture}
\; \rightarrow \; \includegraphics[width=0.005\textheight]{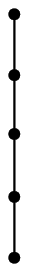}  
\begin{picture}(0,0)
\put(1,46){\footnotesize{$1$}}
\put(1,34){\footnotesize{$2$}}
\put(1,22){\footnotesize{$3$}}
\put(1,11){\footnotesize{$4$}}
\put(1,-1){\footnotesize{$5$}}
\end{picture}
\qquad ; \quad \ldots \\
\nonumber
\end{eqnarray}
Indeed, for \emph{every} n-point function the same cancellations seen in the examples above occur, causing the sum of all the graphs generated from the r-functions and the c-functions to vanish, with the only exception of ladder graphs.
Let us partition the graphs generated by the graphical procedure into three classes: r-graphs with a stalk, r-graphs without a stalk and c-graphs. In the figure below we report the example of the 5-point correlation function:
\begin{eqnarray} \label{groups}
& & \Big( \; \includegraphics[width=0.05\textwidth]{bt5_1.eps} \; + \; \includegraphics[width=0.06\textwidth]{bt5_9.eps} \;  - \; 2 \; \includegraphics[width=0.05\textwidth]{bt5_3.eps} \;  - \; 2 \; \includegraphics[width=0.05\textwidth]{bt5_5.eps} \;  + \; 6 \; \includegraphics[width=0.05\textwidth]{bt5_7.eps} \; \Big) \; \nonumber \\
\cup \; & & \Big( - \; \includegraphics[width=0.05\textwidth]{bt5_2.eps} \; - \; \includegraphics[width=0.06\textwidth]{bt5_10.eps} \; + \; 2 \; \includegraphics[width=0.05\textwidth]{bt5_4.eps} \; + \; 2 \; \includegraphics[width=0.05\textwidth]{bt5_6.eps} \; - \; 6 \; \includegraphics[width=0.05\textwidth]{bt5_8.eps} \; \Big) \; \\
\cup \; & & \Big( \; \includegraphics[width=0.05\textwidth]{gc5_1.eps} \; + \; \includegraphics[width=0.07\textwidth]{gc5_2.eps} \; - \; 5 \; \includegraphics[width=0.06\textwidth]{gc5_3.eps} \; - \; 2 \; \includegraphics[width=0.05\textwidth]{gc5_4.eps} \; + \; 26 \; \includegraphics[width=0.05\textwidth]{gc5_5.eps} \; \Big) \nonumber
\end{eqnarray}
It can be proven that r-graphs without a stalk completely cancel the c-graphs (\ref{rpetc}) and that r-graphs with a stalk cancel one another with the exception of the ladder graph (\ref{rpet}). These facts lead us to the following grand result, sealing the identity between the particle (\ref{nptk}) and field-theoretic approaches:
\begin{equation} \label{grand}
\fl
\langle \varrho(k_1,t_1)\dots\varrho(k_n,t_n) \rangle_c \; = \; \includegraphics[width=0.025\textwidth]{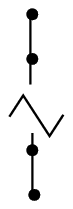} 
\begin{picture}(0,0)
\put(-3,34){\footnotesize{$1$}}
\put(-3,24){\footnotesize{$2$}}
\put(-3,0){\footnotesize{$n$}}
\end{picture}
\; = \; \rho_0 \delta(k_1+\ldots+k_n) \exp \Big( T\sum_{i<j}^n k_i \cdot k_j |t_i-t_j| \Big)
\end{equation}
where the last term has been complemented with a factor $\rho_0 (2\pi)^d \delta(k_1+\ldots+k_n)$, $\rho_0$ coming from the decomposition (\ref{prop-dec}) and the $\delta$ originating from the global conservation of momentum associated to the Feynman diagrams.

\section{The Poissonian distribution and a remark on coarse-graining} \label{poiss}

After Fourier-transforming (\ref{grand}), the analytic expression for an n-point density correlation function in direct space is identical to (\ref{npt-part}): it follows then that the equal-time limit of an n-point function is $\rho_0 \delta(x_1-x_n)\ldots \delta(x_{n-1}-x_n)$. At equal times the density field is thus delta-correlated in space, as expected for a non-relativistic theory which is both causal and local. \\
\\
Let us consider the probability distribution function (PDF) for the equilibrium density measured at a single point in space-time: the equal-time limit of an n-point function is the n$^{th}$ moment of such a PDF. Since a probability distribution is uniquely determined by its moments and we know from (\ref{grand}) that all the cumulants are equal to $\rho_0$, it follows immediately that the PDF of the local density is a Poissonian distribution  with average equal to $\rho_0$.
Whereas this statement is obvious when considered from a particle viewpoint, it is highly non-trivial from a field-theoretic perspective, involving the resummation of an infinite class of Feynman diagrams. \\
\\
For an infinite system, $\lambda = \rho_0^{-1/d}$ is the only natural length-scale. In principle we may want to introduce by hand a second length-scale $l$, which could be a standard unit of length (e.g.\ 1 cm) or being imposed by the spatial resolution of an experiment. Then, $\rho_0$ can be considered as a free parameter.
In the large $\rho_0$ limit the Poisson distribution is well approximated by a Gaussian with average $\rho_0$ and standard deviation $\sigma=\sqrt{\rho_0}$. More precisely, when measured at macroscopic space-time intervals, the 1-point function is $\rho_0 \gg 1$, the 2-point functions are $\mathcal{O}(1)$ and all higher n-point function are suppressed, being $\mathcal{O}(1/\rho_0^{n-2})$. Hence, the effective behavior of the system can be approximated by a free field theory. \\
Yet, for $\rho_0\lesssim 1$, the Poisson distribution is markedly different from a Gaussian\footnote{No inter-particle potential would ever give rise to a genuine free field theory: in that case, in fact, the PDF for the density would be a Gaussian and the density field would not be positive definite.}: the hypothetical experiment is probing the system down to length-scales at which the granularity of the gas is manifest.
The vertex appearing in the field theory, which faithfully reproduces the Poissonian statistics, is then the vestige of the particle nature of the gas.
Its relevance is crucial at low densities where the free particle system is described by a strongly interacting field theory.
Note that the length-scale $\lambda$ merely identifies the cross-over between ``free'' and interacting theory but by no means it represents a hard cutoff scale for the system. Again, any coarse-graining procedure would be purely arbitrary, the field theory being well-defined at any length-scale.

\section{Time-reversal symmetry and fluctuation-dissipation theorem} \label{symm}

This last section contains a brief review of the time-reversal symmetry of the action and the fluctuation-dissipation relation it implies. Also, we will finally focus on the mechanism through which the symmetry is enforced at the level of correlation functions. \\
\\
The path integral (\ref{genfunc}) and the action (\ref{action}) define the field theory for the Brownian gas.
The corresponding equation of motion for the physical field $\varrho$ is Dean's equation (\ref{dean-eq}) and it can be interpreted as a continuity equation for the density:
\begin{displaymath}
\partial_t \varrho =  \nabla \cdot j \qquad \textrm{with current:} \qquad j = -\xi\sqrt{\varrho} - T \nabla \varrho
\end{displaymath}

\subsection{Response functions} \label{symm1}

An external force $f(x)$ acting on the density field shifts the current:
\begin{displaymath}
j(x) \quad \to \quad j(x) + \rho(x) f(x) \quad = \quad j(x) - \rho(x) \nabla \mu(x)
\end{displaymath}
the last equality being meaningful only if $f$ is exact and a scalar potential $\mu$ can be defined: $f(x)=-\nabla \mu(x)$.
Dean's equation can then be written as:
\begin{eqnarray}
\partial_t \varrho &=& \nabla \cdot \left( \xi\sqrt{\varrho} \right) + T \nabla^2 \varrho + \nabla \cdot (\varrho\nabla\mu) \nonumber \\
&=&  \nabla \cdot \left( \xi\sqrt{\varrho} \right) + \nabla \cdot \left( \varrho \nabla \frac{\delta F}{\delta \varrho} \right)
\nonumber
\end{eqnarray}
where the free energy functional $F$ is defined as:
\begin{displaymath}
F[\varrho] = \int \mathrm{d}^dx \; \Bigg[ T \varrho(x,t) \Big( \ln \frac{\varrho(x,t)}{\rho_0} - 1 \Big) + \mu(x,t) \varrho(x,t)  \Bigg]
\end{displaymath}
The physical interpretation of $\mu$ as a chemical potential field linearly coupled to the density is now fully transparent. Notably, however, in the field theory $\mu$ appears in a term of the MSR action which is not quadratic in the fields:
\begin{eqnarray} \label{actionmu}
\fl
S = \int \mathrm{d}^dx \, \mathrm{d}t \;  \Big\{ i\hat{\phi}(x,t)[\partial_t \varrho(x,t)-T\nabla^2\varrho(x,t) -\nabla \cdot (\varrho\nabla\mu)] + T\varrho(x,t)[\nabla i\hat{\phi}(x,t)]^2 \Big\} \\
\nonumber
\end{eqnarray}
We are now interested in the linear response of a generic local functional of the density $\varrho$ to the external field $\mu$, which is defined as:
\begin{displaymath}
R_A(x,x';t,t') \equiv \left. \left\langle \frac{\delta A[\varrho](x,t)}{\delta \mu(x',t')} \right\rangle_{\mu} \;
\right|_{\mu=0}
\end{displaymath}
Integrating (\ref{actionmu}) by parts and taking the functional derivative of the generating functional with respect to $\mu$ yields the field-theoretic expectation value:
\begin{equation} \label{response}
R_A(x,x';t,t') = \left\langle A[\varrho](x,t) \; \nabla \cdot [ (\nabla i\hat\phi ) \varrho ](x',t') \right\rangle
\end{equation}
The particular case $A[\varrho]=\varrho$ makes explicit the distinction mentioned at the end of Section \ref{Feynrules} between the actual response function $\langle \rho(x,t) \nabla \cdot [ (\nabla i\hat\phi) \varrho](x',t') \rangle$ and the pseudo-response functions given by the propagators $\langle i\hat{\phi}(x',t')\rho(x,t) \rangle$ and $\langle \rho(x,t) i\hat{\phi}(x',t') \rangle$.
We now show how a symmetry of the action implies the fluctuation-dissipation theorem involving the proper response function defined in (\ref{response}).

\subsection{Symmetry of the action}

The action (\ref{action}) is invariant, modulo total derivatives, under the time-reversal transformation $\mathcal{T}$, implemented as:
\begin{equation} \label{symmx}
\mathcal{T}: \left\{ \begin{array}{l @{\quad \rightarrow \quad} l}
t & \tau = -t \nonumber \\
\rho(x,t) & \rho(x,\tau) \nonumber \\
\partial_t \rho(x,t) & -\partial_{\tau} \rho(x,\tau) \nonumber \\
i\hat{\phi}(x,t) & -i\hat{\phi}(x,\tau) - \ln{\frac{\varrho(x,\tau)}{\rho_0}} \nonumber
\end{array} \right.
\end{equation}
In general, the invariance of the action under a field transformation is not sufficient to guarantee that such a transformation is a symmetry: in fact, a non-linear field redefinition might lead to a non-trivial Jacobian. We now argue that, although the transformation (\ref{symmx}) is non-linear, its Jacobian is just a simple constant. Consider a finite system with discretized space: a field configuration is then defined by a $2N$-vector $v=(i\hat{\phi}_1,\ldots,i\hat{\phi}_N;\rho_1,\ldots,\rho_N)$ and the Jacobian matrix $J$ associated to (\ref{symmx}) is:
\begin{displaymath}
J_{i,j} = \frac{\partial (\mathcal{T} v)_i}{\partial v_j} = \left( \begin{array}{c|c} A & B \\ \hline C & D \end{array} \right)
\end{displaymath}
with:
\begin{displaymath}
\fl
\qquad \quad A=-\mathbb{I}_{N\times N} \quad ; \quad B=-\left( \begin{array}{c c c} \rho_1^{-1} & 0 & 0 \\ 0 & \ddots & 0 \\ 0 & 0 & \rho_N^{-1} \end{array} \right) \quad ; \quad C=0_{N\times N} \quad ; \quad D=\mathbb{I}_{N\times N}
\end{displaymath}
Since $J$ is block triangular, $|\det(J)|=|\det(A)\det(D)|=|\pm 1|$ and (\ref{symmx}) is indeed a symmetry of the field theory. \\
\\
The study of correlation functions is easier in Fourier space. The action (\ref{action}) can then be rewritten as:
\begin{eqnarray}
S = \int_{k,t} \Bigg\{ \Big[ i\hat{\phi}(k,t) \partial_t \rho(-k,t) \Big] &-& \Big[ i\hat{\phi}(k,t) \int_{k^{\prime}} \frac{1}{T} Q(k,k^{\prime},t) \frac{\delta F}{\delta \rho(k^{\prime},t)} \Big] \nonumber \\
&-& \Big[ i\hat{\phi}(k,t) \int_{k^{\prime}} Q(k,k^{\prime},t) i\hat{\phi}(k^{\prime},t) \Big] \Bigg\} \nonumber
\end{eqnarray}
where: $\int_k \equiv \int \frac{\mathrm{d}^dk}{(2\pi)^d}$ and $\int_t \equiv \int \mathrm{d}t$.
The integral kernel $Q$ is defined as:
\begin{equation} \label{Q}
Q(k,k^{\prime},t) = T k\cdot k^{\prime} \rho(-k-k^{\prime},t)
\end{equation}
and $Q^{-1}$ is such that:
\begin{displaymath}
\int_{k^{\prime}} Q(k,k^{\prime},t) Q^{-1}(k^{\prime},k^{\prime\prime},t) = \delta (k-k^{\prime\prime})
\end{displaymath}
The time-reversal symmetry can now be implemented as:
\begin{displaymath}
\mathcal{T}: \left\{ \begin{array}{l @{\quad \rightarrow \quad} l}
t & \tau = -t \nonumber \\
\rho(k,t) & \rho(k,\tau) \nonumber \\
\partial_t \rho(k,t) & -\partial_{\tau} \rho(k,\tau) \nonumber \\
i\hat{\phi}(k,t) & -i\hat{\phi}(k,\tau) - \int_{k^{\prime\prime}} Q^{-1}(k,k^{\prime\prime},t) \partial_{\tau} \rho(-k^{\prime\prime},\tau) \nonumber 
\end{array} \right. \nonumber
\end{displaymath}
where the last line can also be expressed as:
\begin{displaymath}
\int_{k^{\prime}} Q(k,k^{\prime},t) i\hat{\phi}(k^{\prime},t) \quad \rightarrow \quad - \int_{k^{\prime}} Q(k,k^{\prime},t) i\hat{\phi}(k^{\prime},\tau) - \partial_{\tau} \rho(-k,\tau) 
\end{displaymath}
Notice that $\int_{k^{\prime}} Q(k,k^{\prime},t) i\hat{\phi}(k^{\prime},t)$ is nothing but the Fourier-transformed version of the composite operator $\nabla \cdot [ (\nabla i\hat\phi) \varrho ]$ appearing in (\ref{response}).
The FDT associated to any observable A is then:
\begin{eqnarray} \label{finalfdt}
&& \int_{k^{\prime}} \langle A(k,t) Q(k,k^{\prime},t^{\prime}) i\hat{\phi}(k^{\prime},t^{\prime}) \rangle \stackrel{\mathcal{T}}{\longrightarrow} \\ 
&& \stackrel{\mathcal{T}}{\longrightarrow} - \int_{k^{\prime}} \langle A(k,\tau) Q(k,k^{\prime},\tau^{\prime}) i\hat{\phi}(k^{\prime},\tau^{\prime}) \rangle - \partial_{\tau^{\prime}} \langle A(k,\tau) \rho(-k,\tau^{\prime}) \rangle \nonumber
\end{eqnarray}
In particular, we can consider $A\equiv\rho$ and $t^{\prime} < t$ so that one of the response terms vanishes because of causality. Assuming time-translation and time-reflection invariance, properties which are certainly true at equilibrium, we finally obtain the FDT:
\begin{equation} \label{finalsimplefdt}
\int_{k^{\prime}} T k\cdot k^{\prime} \langle \rho(k,t) \rho(-k-k^{\prime},t^{\prime}) i\hat{\phi}(k^{\prime},t^{\prime}) \rangle = \partial_{t^{\prime}} \langle \rho(k,t) \rho(-k,t^{\prime}) \rangle
\end{equation}

\noindent \\
A few general properties useful in the following computations are:
\begin{itemize}
\item[-] All the FDT's written above hold only once the sum of connected \emph{and disconnected} correlators is considered.
\item[-] For the gas, any correlator containing more than one $i\hat{\phi}$ field vanishes.
\item[-] $\langle i\hat{\phi} \rangle \equiv 0$.
\item[-] Each connected n-point function (n$>2$) involving a field $i\hat{\phi}$ evaluated at the same time as another field $\rho$ is identically zero. This can be easily argued from the integration domains in (\ref{R3}), (\ref{R4}), (\ref{R5}) which vanish if $m_{2\dots n}=t_1$ whereas the integrand is a continuous function.
\end{itemize}
Verifying (\ref{finalsimplefdt}) with $A\equiv\rho$ using the Feynman rules for the propagators is a simple exercise, yielding:
\begin{equation} \label{simpleFDT}
\fl
-T k^2 \rho_0 e^{-Tk^2(t-t^{\prime})} \theta(t-t^{\prime}) = +T k^2 \rho_0 e^{-Tk^2(t^{\prime}-t)} \theta(t^{\prime}-t) - Tk^2 \rho_0 e^{-Tk^2|t-t^{\prime}|} \mathrm{sgn}(t^{\prime}-t) \nonumber
\end{equation}

\noindent \\
When computing explicitly (\ref{finalfdt}) in the case $A\equiv\rho\rho$ or higher powers of $\rho$ interesting cancellations occur between different disconnected correlators. Let us consider the simplest case (assuming $t_1<t_2<t_3$):
\begin{displaymath}
\fl
\int_{k^{\prime}} \langle \rho(k_3,t_3) \rho(k_2,t_2) Q(k_2+k_3,k^{\prime},t_1) i\hat{\phi}(k^{\prime},t_1) \rangle = \partial_{\tau_1} \langle \rho(k_3,\tau_3) \rho(k_2,\tau_2) \rho(-k_3-k_4,\tau_1)  \rangle
\end{displaymath}
After substituting (\ref{Q}) and expanding in the non-vanishing disconnected parts, the formula above can be written in short-hand notation as:
\begin{eqnarray} \label{shorth}
&& \int_{k^{\prime}} T (k_2+k_3)\cdot k^{\prime} \Big( \langle \rho_3 \rho_2 i\hat{\phi}_1 \rangle_c \langle \rho_1 \rangle + \langle \rho_3 i\hat{\phi}_1 \rangle_c \langle \rho_2 \rho_1 \rangle_c + \nonumber \\
&& \langle \rho_2 i\hat{\phi}_1 \rangle_c \langle \rho_3 \rho_1 \rangle_c + \langle \rho_3 i\hat{\phi}_1 \rangle_c \langle \rho_2 \rangle \langle \rho_1 \rangle + \langle \rho_2 i\hat{\phi}_1 \rangle_c \langle \rho_3 \rangle \langle \rho_1 \rangle \Big) = \\
&& = \partial_{\tau_1} \Big( \langle \rho_3 \rho_2 \rho_1 \rangle_c + \langle \rho_3 \rho_1 \rangle_c \langle \rho_2 \rangle + \langle \rho_2 \rho_1 \rangle_c \langle \rho_3 \rangle \Big) \nonumber
\end{eqnarray}
In particular we now show that the following equality holds:
\begin{equation} \label{note}
\fl
\int_{k^{\prime}} T (k_2+k_3)\cdot k^{\prime} \Big( \langle \rho_3 \rho_2 i\hat{\phi}_1 \rangle_c \langle \rho_1 \rangle + \langle \rho_3 i\hat{\phi}_1 \rangle_c \langle \rho_2 \rho_1 \rangle_c + \langle \rho_2 i\hat{\phi}_1 \rangle_c \langle \rho_3 \rho_1 \rangle_c \Big) = \partial_{\tau_1} \langle \rho_3 \rho_2 \rho_1 \rangle_c
\end{equation}
The remaining terms appearing in (\ref{shorth}) and not in (\ref{note}) contain 1 and 2-point functions only and are easily proven to be equal using (\ref{simpleFDT}).
A key observation is that conservation of momentum implies different values for $k^{\prime}$ in the l.h.s.\ of (\ref{note}): $k^{\prime}=-k_2-k_3$, $k^{\prime}=-k_3$ and $k^{\prime}=-k_2$ respectively. Then, graphically:
\begin{eqnarray} \label{fdt3}
&& T \rho_0 \Big[ - (k_2+k_3)\cdot (k_2+k_3) \Big( \; \includegraphics[width=0.04\textwidth]{bt3_1.eps} \begin{picture}(0,0)
\put(-7,23){\footnotesize{$1$}}
\put(-5,11){\footnotesize{$2$}}
\put(-20,-7){\footnotesize{$2$}}
\put(-3,-7){\footnotesize{$3$}}
\end{picture} \; - \; \includegraphics[width=0.04\textwidth]{bt3_2.eps}
\begin{picture}(0,0)
\put(-6,11){\footnotesize{$1$}}
\put(-20,-7){\footnotesize{$2$}}
\put(-3,-7){\footnotesize{$3$}}
\end{picture} \; \Big) \\
&& - (k_2+k_3)\cdot k_3 \; \includegraphics[width=0.04\textwidth]{bt3_2.eps} \begin{picture}(0,0)
\put(-6,11){\footnotesize{$1$}}
\put(-20,-7){\footnotesize{$2$}}
\put(-3,-7){\footnotesize{$3$}}
\end{picture} \; - (k_2+k_3)\cdot k_2 \; \includegraphics[width=0.04\textwidth]{bt3_2.eps} \begin{picture}(0,0)
\put(-6,11){\footnotesize{$1$}}
\put(-20,-7){\footnotesize{$2$}}
\put(-3,-7){\footnotesize{$3$}}
\end{picture} \;\, \Big] \; = \; \rho_0 \; \partial_{\tau_1} \; \includegraphics[width=0.007\textheight]{ladder3.eps}
\begin{picture}(0,0)
\put(0,27){\footnotesize{$1$}}
\put(0,13){\footnotesize{$2$}}
\put(0,0){\footnotesize{$3$}}
\end{picture} \nonumber
\end{eqnarray}
which is clearly true after simplifying the wedge graphs and considering the graphical rules of section \ref{graphrule}. There are two points worth noting: first of all, the non-linear symmetry leads to cross-cancellations between disconnected correlation functions of different rank. Second, the particular dot-product structure of the vertex, together with conservation of momentum, once again produce interesting combinatorial factors which, in the particular case of (\ref{fdt3}), allow a \emph{single} wedge graph to cancel \emph{two} wedge graphs. A generalization of this kind of computations is not a hard exercise.

\section{Conclusions and outlook}

As a complementary approach to the direct development of approximations for an interacting fluid, in this paper we have analyzed a non-interacting gas of Brownian particles; the system has been fully characterized with the computation of all the n-point functions.
The exact correspondence between the particle and the field-theoretic formalism has been established by the non-perturbative calculation of the ground state of the field theory and the determination of the Poissonian probability distribution for the density field.
A non-perturbative treatment of the gas-vertex is thus necessary not only to preserve the fluctuation-dissipation theorem in bare perturbation theory, as already argued in previous works \cite{ABL06}, but also, and more fundamentally, to define the unperturbed ground state as the starting point for a perturbative expansion of physical observables in the inter-particle potential. \\
\\
At the fundamental level, we have shown that the system can be described \emph{exactly} by a field theory and that no coarse-graining procedure is required to define the density field appearing in the path integral formulation. Hence, when considering the interacting theory, any divergence arising from the computation of loop diagrams cannot be ascribed to the effective nature of the field-theoretic description: instead it should be considered as a genuine feature of the problem and properly renormalized.
Along the same lines, the particle origin of the field theory, expressed through the vertex, might determine and naturally constrain the renormalizability properties of the interacting theory. \\
\\
Finally, the pattern through which the FDT is verified on the correlation functions of the gas could shed light on its effect on the combinatorics of the interacting case, where the interest is in the study of high order correlators to define a diverging dynamical length scale \cite{BB04}.

\ack
We warmly thank K. Barros, G. Biroli, P. Krapivsky and K.Yeats for useful discussions. AV has been supported by NSF grants DMS-0603781 and DMR-0403997. LFC is a member of Institut Universitaire de France.

\appendix

\section{Definitions and conventions} \label{defconv}

\subsubsection*{Fourier Transforms}

For a finite system, in standard notation and using $\delta_k$ as a shorthand notation for $\delta_{k,0}$:
\begin{eqnarray}
f_k(t) = \frac{1}{V}\int_V d^dx \; e^{-ik\cdot x}f(x,t) \qquad &;& \qquad f(x,t) = \sum_k e^{ik\cdot x} f_k(t) \nonumber \\
\sum_k e^{ik\cdot x} = V \delta(x) \qquad &;& \qquad \int_V d^dx \; e^{-ik\cdot x} = V \delta_k \nonumber
\end{eqnarray}
In the field-theoretic formalism there is no need to consider a finite number of particles and a finite volume as an intermediate step, the density being the only meaningful quantity. The formulae above are then modified as follows:
\begin{eqnarray}
f(k,t) = \int d^dx \; e^{-ik\cdot x}f(x,t) \qquad &;& \qquad f(x,t) = \int \frac{d^dk}{(2\pi)^d} e^{ik\cdot x} f(k,t) \nonumber \\
\int \frac{d^dk}{(2\pi)^d} e^{ik\cdot x} = \delta(x) \qquad &;& \qquad \int d^dx \; e^{-ik\cdot x} = (2\pi)^d \delta(k) \nonumber
\end{eqnarray}
where all the integrations extends over the entire $d$-dimensional space. \\
\\
Note that due to the different definitions for the Fourier transforms, the dimensions of the density field defined in (\ref{densfield}) are different in Fourier space between the particle and the field-theoretic formalism. For particle systems: $\varrho_{k=0}(t)=\frac{1}{V} \int_V d^dx \; \varrho(x,t) = \rho_0$ with dimensions of volume$^{-1}$; in the field-theoretic formulation: $\varrho(k=0,t)=\int d^dx \; \varrho(x,t) = N$ which is dimensionless.

\subsubsection*{Averages and a useful identity}

\begin{itemize}
\item[-] $\langle A \rangle_{\eta}$ indicates the average of the noise-dependent observable $A$ over the realizations of the noises $\eta_1, \ldots, \eta_N$.
\item[-] $\langle A \rangle_{I.P.}$ indicates instead the average over the initial positions of the particles. It is defined in such a way that $\langle 1 \rangle_{I.P.} = 1$:
\begin{displaymath}
\langle A \rangle_{I.P.} \qquad := \qquad \frac{1}{V} \int_V d^dx_1(0) \ldots \frac{1}{V} \int_V d^dx_N(0) \; A
\end{displaymath}
\end{itemize}
In the particle formalism, with the above conventions for the Fourier transforms, identity (\ref{averageIP}) follows as a straightforward result:
\begin{eqnarray} \label{averageIP}
& &\langle e^{-ik_{j_1}\cdot x_{j_1}(0)}\dots e^{-ik_{j_n}\cdot x_{j_n}(0)} \rangle_{I.P.} \nonumber \\
& & = \left\{ \begin{array}{ll}
\delta_{k_1+\dots+k_n} & \textrm{if all $j$'s are the same} \\
\delta_{k_1+\dots+k_{a-1}+k_{a+1}+\dots+k_n}\delta_{k_a} & \textrm{if all $j$'s but $j_a$ are the same} \\
\ldots & \ldots \\
\delta_{k_1}...\delta_{k_{a-1}}\delta_{k_{a+1}}...\delta_{k_{b-1}}\delta_{k_{b+1}}...\delta_{k_n}\delta_{k_a+k_b} & \textrm{if all $j$'s differ but $j_a=j_b$} \\
\delta_{k_1}\dots\delta_{k_n} & \textrm{if all $j$'s differ} \\
\end{array} \right.
\end{eqnarray}

\section{Combinatorial coefficients}

\subsection{Factorials} \label{factorial}

\textbf{Theorem:} \emph{the coefficient $\mathcal{R}_l$ (defined in section \ref{comb}) associated to a vertex with $l$ dangling lines is $\mathcal{R}_l=(l-1)!$.} \\
\\
We have verified in the text that the relation is true for some $l$:
$\mathcal{R}_2=1=(2-1)!$, as shown in (\ref{R3}) and in the comment below it. Formula (\ref{norat4}) implies $\mathcal{R}_3=2=(3-1)!$. From section \ref{sec5pt} we also know that $\mathcal{R}_4=6=(4-1)!$. \\
\\
Assuming that the relation is valid for vertices having up to $l$ dangling, we now prove it for a vertex with $l+1$ dangling lines. Let us consider one such a vertex, with the upper edge labeled by $l+1$. 
\begin{eqnarray} \label{DS}
\mathcal{R}_l \quad = \quad \quad \includegraphics[width=0.07\textwidth]{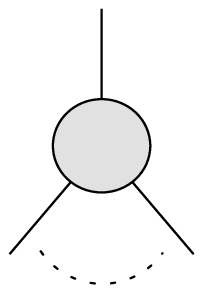} \quad &=& \quad \frac{1}{2!} \quad \Bigg( \quad \includegraphics[width=0.10\textwidth]{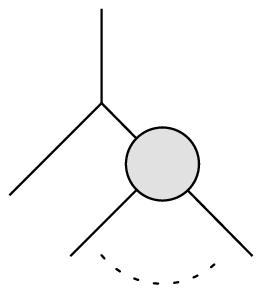} \quad + \quad \includegraphics[width=0.10\textwidth]{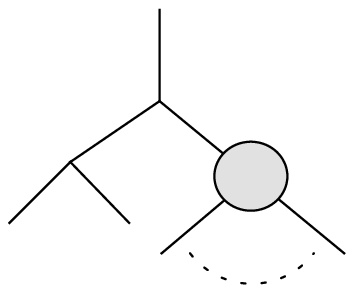} \quad \\
&+& \quad \includegraphics[width=0.12\textwidth]{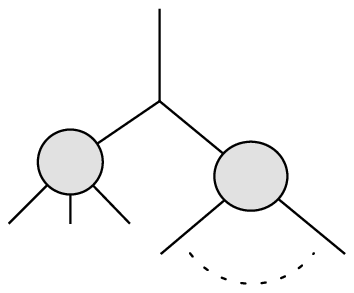} \quad + \quad \dots \quad + \quad \includegraphics[width=0.10\textwidth]{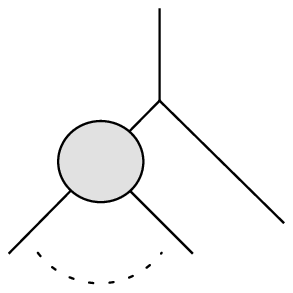} \quad \Bigg) \nonumber
\end{eqnarray}
In a Dyson-Schwinger fashion the earliest vertex is pulled out of the blob.
Conversely, when contracting the edges generated by the extraction of the vertex, the graphs on the r.h.s. generate the graph on the l.h.s..
Note that in (\ref{DS}) any vertex with two dangling lines must be a bare vertex since no loop-graphs contribute to the physical amplitudes. \\
All binary (non-planar) trees are generated by a recursive application of (\ref{DS}). They all appear with coefficient 1 except for one particular case: when $l$ is even, the graph on the r.h.s. for which $r=s=l/2$ appears only once and the factor of $1/2$ out of the braces in (\ref{DS}) is not simplified. However such a graph has a left-right symmetry and because of the very same symmetry a factor 2 will arise in formula (\ref{fact1}) since the partition with $r$ elements and the partition with $s$ elements are equivalent. This cancellation will be tacitly assumed in the argument following (\ref{fact1}). \\
\\
Every term in the r.h.s of (\ref{DS}) represents a partition of the $l$ dangling lines into two groups with $r$ and $s$ dangling lines each, with the constraint $l=r+s$. For a generic partition, the Feynman rule associated to the extracted vertex reads:
\begin{equation} \label{eq}
- \sum_{TI} 2 (k_1+\ldots +k_r)\cdot (k_{r+1}+\ldots +k_l)
\end{equation} 
The $\sum_{TI}$ is over the ${l \choose r}$ possible partitions depicted in (\ref{DS}). Applying conservation of momentum, we can rewrite (\ref{eq}) as:
\begin{eqnarray} \label{fact1}
& & 2 \sum_{TI} (k_1+\ldots +k_r)\cdot (k_{l+1}+k_1+\ldots +k_r) = \\
&& = 2 \sum_{TI} \Big[ (k_1^2+\ldots +k_r^2) + (\sum_{i\neq j}^r k_i \cdot k_j) + k_{l+1}\cdot (k_1+\ldots +k_r) \Big] \nonumber
\end{eqnarray}
The sum over all TI labelings can be easily computed by the following argument: in the sum each $k_i^2$ appears ${l-1 \choose r-1}$ times, since once $k_i$ is fixed to be in the group of $r$ elements, the other $r-1$ elements can be chosen among $l-1$ labels. Similarly, each ordered pair $k_i k_j$ in the inner sum appears ${l-2 \choose r-2}$ times, since once $k_i$ and $k_j$ are fixed to be in the group of $r$ elements, the other $r-2$ elements can be chosen among $l-2$ labels.
\begin{eqnarray}
\fl
&& =  {l-1 \choose r-1}(k_1^2+... +k_l^2) + {l-2 \choose r-2}\sum_{i\neq j}^l k_i \cdot k_j + {l-1 \choose r-1} k_{l+1}\cdot (k_1+...+k_l)
\end{eqnarray}
\begin{eqnarray}
&& =  {l-1 \choose r-1}(k_1^2+\ldots +k_l^2) + {l-2 \choose r-2}\sum_{i=1}^l k_i \cdot (-k_i -k_{l+1}) - {l-1 \choose r-1} k_{l+1}^2 \nonumber \\
&& =  {l-1 \choose r-1}(-k_{l+1}^2+k_1^2+\ldots +k_l^2) + {l-2 \choose r-2}\sum_{i=1}^l (-k_i^2 + k_{l+1}^2) \nonumber \\
&& = {l-2 \choose r-1}(-k_{l+1}^2+k_1^2+\ldots +k_l^2) \nonumber
\end{eqnarray} 
where in the second and third steps, conservation of momentum has been used. \\
Summing over all possible partitions of labels into two blobs and associating to each blob with $n$ dangling lines the appropriate $\mathcal{R}_n=(n-1)!$, we finally obtain:
\begin{equation}
\sum_{r=1}^{l-1} {l-2 \choose r-1}(r-1)!(l-r-1)! \; = \; (l-2)! \sum_{r=1}^{l-1} 1 \; = \; (l-1)!
\nonumber
\end{equation}

\subsection{Generalized Stirling numbers} \label{stirling}

\textbf{Theorem:} \emph{the coefficient $\mathcal{C}_n$ (defined in section \ref{goto}) associated to the c-vertex with $n$ dangling lines appearing in the computation of c-functions is: $\mathcal{C}_n=S_{n-2}$, where $S_{n}$ is the sequence of generalized Stirling numbers (Sloane A001705, \cite{OEIS}).} \\
\\
We can write a Dyson-Schwinger equation similar to (\ref{DS}), with the important difference that the c-vertex needs to have 3 dangling blobs: the first blob with $l$ dangling lines, the second one with $m$, and the third one with $n-l-m$. The factor $1/2!$ in (\ref{DS}) becomes $1/3!$ taking into account the possible permutations of blobs. The rigorous argument about the cancellation of symmetry factors for symmetric diagrams is analogous to the one which followed equation (\ref{DS}). Formula (\ref{eq}) is replaced by the Feynman rule for the c-vertex obtained in (\ref{c-vertex}):
\begin{equation} \label{stirl1}
\frac{1}{3!} \sum_{TI} \Big[ (k_1+... +k_l)^2 + (k_{l+1}+...+k_{l+m})^2 + (k_{l+m+1}+...+k_n)^2 \Big]
\end{equation}
In the sum over TI relabelings each $k_i^2$ appears once for each of the possible ${n \choose l} {n-l \choose m}$ labelings, since $k_i$ has to be in one of the three groups in (\ref{stirl1}). 
If we ask an ordered pair $k_i k_j$ to appear in the blob with $l$ dangling lines, this happens ${n-2 \choose l-2} {(n-2)-(l-2) \choose m}$ times, since once $k_i$ and $k_j$ are fixed to be in the group of $l$ elements, the other $l-2$ labels can be chosen among $n-2$; the other labels are then partitioned into blobs with $m$ and $n-l-m$ dangling lines respectively. 
Focusing instead on the blobs with $m$ and $n-l-m$ a given ordered pair appears ${n-2 \choose m-2} {n-m \choose l}$ and ${n-2 \choose l-2} {(n-2)-(l-2) \choose m}$ times respectively.
Finally, to each blob with $r$ dangling lines we associate a factor $(r-1)!$ as derived in the previous section.
The sum over TI labelings can now be written as:
\begin{eqnarray}
& & \frac{1}{3!} \sum_{l=1}^{n-2} \sum_{m=1}^{n-l-1} (l-1)!(m-1)!(n-l-m-1)! \cdot \nonumber \\
& & \cdot \Bigg\{ {n \choose l} {n-l \choose m} \sum_{i=1}^n k_i^2 + \Bigg[ {n-2 \choose l-2} {n-l \choose m} + {n-2 \choose m-2} {n-m \choose l} \nonumber \\
&& + {n-2 \choose n-l-m-2} {l+m \choose l} \Bigg] \sum_{i\neq j}^n k_i\cdot k_j  \Bigg\}
\end{eqnarray}
Exploiting once more conservation of momentum: $\sum_{i\neq j}^n k_i \cdot k_j = -\sum_{i=1}^n k_i^2$. We can then factor out $\sum_{i=1}^n k_i^2$ which always cancels the denominator of the rational function generated by the c-vertex and will be dropped in the following. Simplifying the binomial coefficients we obtain:
\begin{displaymath}
\frac{1}{3!} \sum_{l=1}^{n-2} \sum_{m=1}^{n-l-1} (n-2)! \Big( \frac{n-l}{m(n-l-m)} + \frac{n-m}{l(n-l-m)} + \frac{l+m}{l m} \Big)
\end{displaymath} 
Moreover, the three terms in parentheses clearly give the same result when summed over $l$ and $m$: for simplicity we focus only on the third one and multiply by 3. We thus have, exchanging the order of the summations in the second term:
\begin{eqnarray}
&& \frac{1}{2!} (n-2)! \Big( \sum_{l=1}^{n-2} \sum_{m=1}^{n-l-1} \frac{l}{m} + \sum_{m=1}^{n-2} \sum_{l=1}^{n-m-1} \frac{m}{l} \Big) \nonumber \\
= && \frac{1}{2!} (n-2)! \Big( \sum_{l=1}^{n-2} H_{n-l-1} + \sum_{m=1}^{n-2} H_{n-m-1} \Big) \; = \; S_{n-2} \nonumber
\end{eqnarray}
where by $H_{n}$ we have indicated the harmonic numbers.

\section{Graphology}

\subsection{Correlation functions are represented by ladder graphs} \label{ladders}

\textbf{Theorem:} \emph{the analytic expression associated to an n-point ladder graph, namely:}
\begin{equation} \label{corr-ladder1}
\exp \Big[ -k_1^2(t_2-t_1) - \ldots - (k_1+\ldots+k_{n-1})^2(t_n-t_{n-1}) \Big]
\end{equation}
\emph{is the same as the analytic formula (\ref{nptk}) for the n-point connected correlation function computed in the particle formalism:}
\begin{equation} \label{corr-part}
\langle \varrho(k_1,t_1)\dots\varrho(k_n,t_n) \rangle_{\eta} \quad = \quad \exp \Big( T\sum_{i<j}^n k_i \cdot k_j (t_j-t_i) \Big)
\end{equation}
Examples of ladder graphs are drawn in figure (\ref{nladders}), and the absolute value of the time differences in (\ref{corr-part}) has been removed considering the convention $t_1 < \ldots < t_n$. \\
\\
Let us define the variables $\hat{k}$'s as $\hat{k}_j=\sum_{i=1}^{j} k_i$. 
Hence we can rewrite (\ref{corr-ladder1}) as:
\begin{equation} \label{corr-ladder2}
\exp{\Big( -\sum_{m=1}^{n-1}\hat{k}_m^2 (t_{m+1}-t_m) \Big)}
\end{equation}
We will now prove the theorem, showing that the exponent in (\ref{corr-part}) can be rewritten as the exponent in (\ref{corr-ladder2}). \\
Expanding $(t_j-t_i)$ in a telescopic sum:
\begin{displaymath}
t_j-t_i = (t_j-t_{j-1}) + (t_{j-1}-t_{j-2}) + \dots + (t_{i+1}-t_i) = \sum_{m=i}^{j-1} (t_{m+1}-t_m)
\end{displaymath}
we can write:
\begin{displaymath}
\sum_{j=1}^n \sum_{i=1}^{j-1} k_i \cdot k_j (t_i-t_j) = \sum_{j=1}^n \sum_{i=1}^{j-1} \sum_{m=i}^{j-1} k_i \cdot k_j (t_m-t_{m+1})
\end{displaymath}
Exchanging the order of the summations we obtain:
\begin{equation} \label{1st}
\sum_{j=1}^n \sum_{i=1}^{j-1} \sum_{m=i}^{j-1} \quad \longrightarrow 
\quad \sum_{m=1}^{n-1} \sum_{i=1}^{m} \sum_{j=m+1}^{n} 
\end{equation}
and looking at the innermost summation, conservation of momentum implies:
\begin{equation} \label{2nd}
\Big( \sum_{j=m+1}^{n} k_j \Big) \cdot k_i (t_m-t_{m+1}) = -\sum_{j=1}^{m} k_j \cdot k_i (t_m-t_{m+1})
\end{equation}
Applying (\ref{1st}), (\ref{2nd}) and the definition of $\hat{k}$ it is now easy to pass from the exponent in (\ref{corr-part}) to the exponent in (\ref{corr-ladder2}):
\begin{eqnarray}
& & \sum_{j=1}^n \sum_{i=1}^{j-1} \sum_{m=i}^{j-1} k_i \cdot k_j (t_m-t_{m+1}) = \sum_{m=1}^{n-1} \sum_{i=1}^{m} \sum_{j=m+1}^{n} k_i \cdot k_j
(t_m-t_{m+1}) \\
&& = -\sum_{m=1}^{n-1} \sum_{i=1}^{m} \sum_{j=1}^{m} k_j \cdot k_i (t_m-t_{m+1}) = -\sum_{m=1}^{n-1} \hat{k}_m^2 (t_m-t_{m+1})
\end{eqnarray}

\subsection{Cancellations among r-graphs with a stalk} \label{rpet}

\textbf{Theorem:} \emph{in the sum over topologically inequivalent (TI) labelings, r-graphs with a stalk cancel one another, leaving only one n-ladder graph.} \\
\\
Examples of r-graphs with a stalk appear in the first line of (\ref{groups}).
Let us assume that one of the r-vertices of a graph has $n>2$ dangling lines. The combinatorial factor associated to such vertex is $\mathcal{R}_n=(n-1)!$. In the sum over TI labelings, graphs with such a vertex can be obtained in several ways: either from a labeling of the graph itself or from a labeling of a ``less contracted'' graph which induce one or more contractions generating the vertex. Since $n>2$ we know that at least one contraction has occurred. We can now work backwards and undo the contraction to obtain all the possible graphs which could have generated the original graph. From the group of $n$ labels we can pull out a group of $k$ labels, leaving a vertex with $n-k+1$ dangling lines in such a way that the largest label is in the extracted group. The rules in section \ref{comb} imply then that the edge we have just created must indeed be contracted. The combinatorial factor associated to the two vertices just created are $(k-1)!$ and $(n-k)!$ respectively. Note that since the largest label has to be in the group with $k$ elements, among the $n-1$ labels left we have to choose $k-1$ of them. Summing over all possibilities:
\begin{equation} \label{canc1}
\sum_{k=2}^{n-1} {n-1 \choose k-1}(k-1)!(n-k)! = (n-1)!
\end{equation}
Remember now that each contraction brings a factor of $(-1)$ so that all the graphs we generated with one contraction less than $X$ appear with a relative minus sign. But (\ref{canc1}) implies that the numerical value of the coefficient is the same and the sum over the whole group of graphs vanishes. \\
On the other side ladder diagrams survive. In fact, in order to create a ladder, from a blob with $n$ labels we pull out recursively a group of $n-1$ labels which do \emph{not} contain the earliest time among the $n$ labels. Then, the argument above leading to the graph cancellations does not apply.

\subsection{c-graphs cancel r-graphs without a stalk} \label{rpetc}

\textbf{Theorem:} \emph{in the sum over topologically inequivalent (TI) labelings, c-graphs completely cancel r-graphs without a stalk.} \\
\\
Examples of r-graphs without a stalk and c-graphs appear in the second and third line of (\ref{groups}) respectively. \\
Let us consider an $n+1$ point function. r-graphs without a stalk have $n$ leaves and c-graphs have $n+1$ leaves.
It is useful to note that for r-graphs the label `1' is constrained by causality to appear on the upper vertex (root), as a consequence of the contraction of the stalk. For c-graphs, instead, the c-vertex itself is always labeled by `1' (`1' being the smallest among all the descendant labels), another label `1' has to be placed on some leaf, and because of this some edges need to be contracted according to the rules discussed in section \ref{graphrule}.
The key idea of the proof is that for an r-graph the number of edges dangling from the root is independent of the labeling, whereas for c-graphs edge contractions cause this number to vary. \\
Let us consider an r-graph $X$ whose root has $l$ dangling lines with the associated combinatorial factor $(l-1)!$.
It is possible that particular labelings of some c-graphs $Y$ which initially have $3,\ldots,l+1$ dangling lines produce $X$ after operating the contractions induced by the labelings. Those graphs will be multiplied by the generalized Stirling numbers $S_1,\ldots,S_{l-1}$. In particular, we can construct all the graphs $Y$ starting from a c-vertex with $l+1$ dangling lines and ``pulling down'' groups of edges with the constraints that a c-vertex must have at least three dangling lines and that `1' labels one of the lowest leaves in such a way to induce a contraction leading back to a root with $l$ dangling lines. \\
An example should make things clear. One of the r-graphs without a stalk contributing to the computation of the 5-point function, with the appropriate factor of $(4-1)!=6$, is:
\begin{equation}
X = 6 \quad \includegraphics[width=0.05\textwidth]{bt5_8.eps}
\end{equation}
Among the c-graphs, the ones which follow generate $X$ if, as explained above, `1' labels one of the lowest leaves.
The coefficient in front of each graph is made up of: a sign which keeps track of the number of contracted edges; the generalized Stirling number associated to the c-vertex; the product of coefficients associated to the underlying r-vertices; the number of labelings which leave `1' to label one of the lowest leaves.
\begin{eqnarray}
&& Y's \quad = \quad -26 \cdot 1 \cdot 1 \quad \includegraphics[width=0.05\textwidth]{gc5_5.eps} \quad + \quad 5 \cdot 1 \cdot {4 \choose 1} \quad \includegraphics[width=0.06\textwidth]{gc5_3.eps} \quad \nonumber \\
&& + \quad 1 \cdot 2 \cdot {4 \choose 2} \quad \includegraphics[width=0.05\textwidth]{gc5_4.eps} \quad - \quad 1 \cdot 1 \cdot 1 \cdot {4 \choose 2}{2 \choose 1} \quad \includegraphics[width=0.05\textwidth]{gc5_1.eps} \quad \rightarrow \quad -6 \quad \includegraphics[width=0.05\textwidth]{bt5_8.eps} \nonumber
\end{eqnarray}
or equivalently:
\begin{equation} \label{stirllc}
-S_3 + S_2 {4 \choose 1} + S_1 {4 \choose 2} 2! + S_1 {4 \choose 2}{2 \choose 1} = -(4-1)!
\end{equation}
We see that $(l-1)!$ is exactly canceled by a linear combination of the generalized Stirling numbers $S_1,\ldots,S_{l-1}$. In this case $l=4$. \\
\\
Working out explicitly more complicated graphs through the ``pulling down'' procedure, it is not hard to write the general form for the l.h.s. of (\ref{stirllc}), which is, modulo a sign:
\begin{eqnarray}
&& S_{l-1} \; - \;
\sum_{k_1=1}^{l-2} S_{l-1-k_1} {l \choose k_1} k_1! \; + \;
\sum_{k_2=2}^{l-2} \sum_{k_1=1}^{k_2-1} S_{l-1-k_2} {l \choose k_2} {k_2 \choose k_1} (k_2-k_1)! k_1! \; \nonumber \\
&& - \; \sum_{k_3=3}^{l-2} \sum_{k_2=2}^{k_3-1} \sum_{k_1=1}^{k_2-1} S_{l-1-k_3} {l \choose k_3} {k_3 \choose k_2} {k_2 \choose k_1} (k_3-k_2)! (k_2-k_1)! k_1! \; + \; \ldots \; \nonumber \\
&& = \; S_{l-1} \; - \;
\sum_{k_1=1}^{l-2} S_{l-1-k_1} \frac{l!}{(l-k_1)!} \; + \;
\sum_{k_2=2}^{l-2} S_{l-1-k_2} \frac{l!}{(l-k_2)!} \sum_{k_1=1}^{k_2-1} 1 \; \nonumber \\
&& - \; \sum_{k_3=3}^{l-2} S_{l-1-k_3} \frac{l!}{(l-k_3)!} \sum_{k_2=2}^{k_3-1} \sum_{k_1=1}^{k_2-1} 1 \; + \; \ldots \; \nonumber \\
&& = \; S_{l-1} \; + \; \sum_{j=1}^{\infty} \sum_{k_j=j}^{l-2} (-1)^j S_{l-1-k_j} \frac{l!}{(l-k_j)!} {k_j-1 \choose j-1} \; \nonumber \\
&& = \; S_{l-1} \; + \; \sum_{k=1}^{l-2} S_{l-1-k} \frac{l!}{(l-k)!} \sum_{j=1}^{k} (-1)^j {k-1 \choose j-1} \; \nonumber \\
&& = \; S_{l-1} \; - \; l S_{l-2} \; = \; (n-1)! \nonumber
\end{eqnarray}
\\

\end{document}